\documentclass[11pt]{article}

\usepackage{amsmath,amsthm,amscd,array,amssymb}

\begin{document}

\begin{center}
{\Large{\bf Higher dimensional analogue of the Blau--Thompson model
\\
\medskip\smallskip
and $N_T = 8$, $D = 2$ Hodge--type cohomological gauge theories}}
\\
\bigskip\medskip
{\large{\sc B. Geyer}}$^{a,b}$
\footnote{Email: geyer@itp.uni-leipzig.de}
{\large{\sc and}}
{\large{\sc D. M\"ulsch}}$^{c}$
\footnote{Email: muelsch@informatik.uni-leipzig.de}
\\
\smallskip
{\it $^a$ Universit\"at Leipzig, Naturwissenschaftlich-Theoretisches Zentrum
\\
$~$ and Institut f\"ur Theoretische Physik, D--04109 Leipzig, Germany\\
$^b$ Instituto de Fisica, Universidade de Sao Paulo, 
Sao Paulo, Brasil\\
\smallskip
$\!\!\!\!\!^c$ Wissenschaftszentrum Leipzig e.V., D--04103 Leipzig, Germany}
\\
\bigskip
{\small{\bf Abstract}}
\\
\end{center}

\begin{quotation}
\noindent {\small{The higher dimensional analogue of the Blau--Thompson
model in $D = 5$ is constructed by a $N_T = 1$ topological twist of
$N = 2$, $D = 5$ super Yang--Mills theory. Its dimensional reduction to
$D = 4$ and $D = 3$ gives rise to the B--model and the $N_T = 4$
equivariant extension of the Blau--Thompson model, respectively.
A further dimensional reduction to $D = 2$ provides another example
of a $N_T = 8$ Hodge--type cohomological theory with global symmetry group
$SU(2) \otimes \overline{SU(2)}$. Moreover, it is shown that this theory
possesses actually a larger global symmetry group $SU(4)$ and that it is
agrees with the $N_T = 8$ topological twisting of $N = 16$, $D = 2$ super
Yang--Mills theory.}}
\end{quotation}

\bigskip
\begin{flushleft}
{\large{\bf 1. Introduction}}
\end{flushleft}
\medskip

Some very enlightening, but preliminary attempts have been made to incorporate
into the gauge--fixing procedure of general gauge theories besides the basic
ingredience of the BRST operator $\Omega$ also a co--BRST operator
$^\star \Omega$ which, together with the BRST Laplacian $W$, form the same
kind of superalgebra as the de Rham cohomology operators in differential
geometry  \cite{1}. This allows, according to the Hodge--type decomposition
$\psi = \omega + \Omega \chi + \,^\star \Omega \phi$ of a general quantum
state, by imposing both the BRST condition $\Omega \psi = 0$ and the co--BRST
condition $^\star \Omega \psi = 0$ on $\psi$, to select the uniquely
determined harmonic state $\omega$ thereby projecting onto the subspace of
physical states (for details, see, Section 2 below).

It has been a long--standing problem to present a non--abelian field
theoretical model obeying such a Hodge--type cohomological structure.
Recently, the authors have shown \cite{2} that the dimensional reduced
Blau--Thompson model \cite{3} --- the novel $N_T = 2$ topological twist of
the $N = 4$, $D = 3$ super Yang--Mills theory (SYM) --- gives a prototype
example of a $N_T = 4$, $D = 2$ Hodge--type cohomological gauge theory.
The conjecture, that topological gauge theories could be possible candidates
for Hodge--type cohomological theories was already asserted by van Holten
\cite{4}. In fact, $D = 2$ topological gauge theories \cite{5} are of
particular interest because of their relation to $N = 2$ superconformal
theories \cite{6} and Calabi--Yau moduli spaces \cite{7}.

In the present paper we construct another example of a 2--dimensional
Hodge--type cohomological theory, but now with the largest possible,
$N_T = 8$ topological (co--)shift symmetry and with global symmetry group
$SU(4)$.\footnote{The existence of such a topological theory was already
mentioned in \cite{10}, see, footnote on page 248.}
This is achieved by first introducing a higher
dimensional analogue of the Blau--Thompson model in $D = 5$ by a $N_T = 1$
topological twist of $N = 2$, $D = 5$ SYM with internal symmetry group
$Spin(5) \sim Sp(4)$. The twisting procedure consists simply in taking the
diagonal subgroup of the R--symmetry group $SO(5)$ and the Euclidean rotation
group $SO_E(5)$. The most unusual feature of this topological model is,
analogous to the Blau--Thompson model, that it has no bosonic scalar fields
and hence no underlying equivariant cohomology. We conjecture that this new
topological theory, which localizes onto the moduli space of complexified
flat connections, is the {\it only} one which can be constructed on a
{\it generic} $5$--dimensional Riemannian manifold with $SO(5)$ holonomy.
The other cohomological theories in $D = 5$, which localize onto the
moduli space of instantons, can be obtained by a dimensional reduction
from the higher dimensional analogues of the Donaldson--Witten theory
in $D = 8$ and $D = 7$ \cite{8,9,10}. These theories should be simply
untwisted SYM theories formulated on manifolds with reduced holonomy group
$H \subset SO(5)$.

From this 5--dimensional analogue of the Blau--Thompson model one
gets, by an ordinary dimensional reduction to $D = 4$, the
B--model \cite{11,3}, i.e., one of the 3 inequivalent topological
twists of $N = 4$, $D = 4$ SYM, and by reduction to $D = 3$ the
$N_T = 4$ {\it equivariant extension} of the Blau--Thompson model
\cite{12}. A further dimensional reduction to $D = 2$ leads to a
Hodge--type cohomological theory with global symmetry group $SU(2)
\otimes \overline{SU(2)}$ and $N_T = 8$ scalar supercharges. These
supercharges, in complete analogy to the de Rham cohomology
operators, are interrelated by a discrete Hodge--type $\star$
operation and generate the topological shift and co--shift
symmetries. In accordance with the group theoretical description
of some classes of topologically twisted low--dimensional
supersymmetric world--volume theories \cite{3}, it is shown that
this Hodge--type cohomological theory actually allows for the
larger global symmetry group $SU(4)$ \cite{13}. Moreover, it is
shown that this theory is precisely the topological twisted $N =
16$, $D = 2$ SYM with R--symmetry group $SU(4) \otimes U(1)$. Such
theories are naturally realized as Dirichlet $1$--brane instantons
wrapping around supersymmetric $2$--cycles of Calabi--Yau
$2$--folds (see, e.g., \cite{14,15}).

The paper is organized as follows:
In Sec. 2 we briefly introduce the BRST complex of general gauge theories
and the Hodge--type decomposition.
In Sec. 3 we construct the 5--dimensional analogue of the Blau--Thompson
model by dimensionally reducing $N = 1$, $D = 10$ SYM and performing a
$N_T = 1$ topological twist of the resulting Euclidean $N = 2$, $D = 5$ SYM
with R--symmetry group $SO(5)$.
In Sec. 4 we show that the dimensional reduction of that theory to $D = 4$
and $D = 3$ gives rise to the B--model and the $N_T = 4$ equivariant extension
of the Blau--Thompson model, respectively.
In Sec. 5 we study the invariance properties of the $N_T = 8$ Hodge--type
cohomological gauge theory with global symmetry group $SU(2) \otimes
\overline{SU(2)}$ obtained by a further dimensional reduction to $D = 2$.
In Sec. 6 we show that this theory can be cast into a form with the
larger global symmetry group $SU(4)$.
In Sec. 7 we describe in detail the $N_T = 8$ topological twist of
the Euclidean $N = 16$, $D = 2$ SYM obtained from the $N = 4$, $D = 4$ SYM
via dimensional reduction to $D = 2$, and show that it agrees precisely with
the Hodge--type cohomological theory with global symmetry group $SU(4)$.

\bigskip
\begin{flushleft}
{\large{\bf 2. BRST complex and Hodge--type decomposition}}
\end{flushleft}
\medskip

In this section we give a rough outlet of the BRST complex,
the cohomologies of the (co--) BRST operators and the Hodge--type
decomposition as far as it will be used in this paper.

In order to select uniquely the physical states from the ghost--extended
quantum state space some attempts \cite{1} have been made to
incorporate into the gauge--fixing procedure of general gauge theories
besides the BRST operator $\Omega$ also a co--BRST operator $^\star \Omega$
which, together with the BRST Laplacian $W$, obeys the following
BRST--complex:
\begin{equation}
\label{2.1}
\Omega^2 = 0,
\qquad
^\star \Omega^2 = 0,
\qquad
W = \{ \Omega, \,^\star \Omega \} \neq 0,
\qquad
[ \Omega, W ] = 0,
\qquad
[ \,^\star \Omega, W ] = 0,
\end{equation}
where $\Omega$ and $^\star \Omega$ have opposite ghost number. Obviously,
$^\star \Omega$ can not be identified with the anti--BRST operator
$\bar \Omega$ which anticommutes with $\Omega$.

Representations of this algebra for the first time have been considered by
Nishijima \cite{16}. However, since $\Omega$ and $^\star \Omega$ are
nilpotent hermitian operators they cannot be realized in a Hilbert space.
Instead, the BRST complex has to be represented in a Krein space $\cal K$
\cite{17}. $\cal K$ is obtained from a Hilbert space $\cal H$ with
non--degenerate positive inner product $(\chi , \psi)$ if $\cal H$ will be
endowed also with a self--adjoint metric operator $J \neq 1$, $J^2 = 1$,
allowing for the introduction of another non--degenerate, but indefinite
scalar product $\langle \chi | \psi \rangle := ( \chi, J \psi )$.
With respect to the inner product $\Omega$ and $^\star \Omega =
\pm J \Omega J$ are adjoint to each other, $( \chi , \,^\star \Omega \psi )
= ( \Omega \chi , \psi )$, however they are self--adjoint with respect to the
indefinite scalar product of $\cal K$. Notice, that different inner products
$( \chi , \psi )$ lead to different co--BRST operators!

From these definitions one obtains a remarkable correspondence
between the BRST cohomology and the de Rham cohomology:
\begin{alignat*}{4}
&\hbox{BRST operator}
&\quad&
\Omega,
&\qquad\qquad&
\hbox{differential}
&\quad&
d,
\\
&\hbox{co--BRST operator}
&\quad&
^\star\Omega = \pm J \Omega J,
&\qquad\qquad&
\hbox{co--differential}
&\quad&
\delta = \pm \star d \star,
\\
&\hbox{duality operation}
&\quad&
J,
&\qquad\qquad&
\hbox{Hodge star}
&\quad&
\star,
\\
&\hbox{BRST Laplacian}
&\quad&
W = \{ \Omega, \,^\star\Omega \},
&\qquad\qquad&
\hbox{Laplacian}
&\quad&
\Delta = \{ d, \delta \}.
\end{alignat*}
Because of this correspondence one denotes a state $\psi$
to be BRST (co--)closed iff $\Omega \psi = 0$ ($^\star \Omega \psi = 0$),
BRST (co--)exact iff $\psi = \Omega \chi$ ($\psi = \,^\star \Omega \phi$)
and BRST harmonic iff $W \psi = 0$. Completely analogous to the Hodge
decomposition theorem in differential geometry there exists a corresponding
decomposition of any state  $\psi$ into a harmonic, an exact and a co--exact
state, $\psi = \omega + \Omega \chi + \,^\star \Omega \phi$.
The physical properties of $\psi$ lie entirely within the BRST harmonic part
$\omega$ which is given by the zero modes of $W$; thereby
$W \omega = 0$ implies $\Omega \omega = 0 = \,^\star \Omega \omega$, and
vice versa. The cohomologies of the (co--)BRST operator $\Omega$
(and $^\star\Omega$) are given by equivalence classes of states:
\begin{alignat*}{2}
&{\rm H}(\Omega) = {{\rm Ker}\, \Omega}/{{\rm Im}\, \Omega},
&\qquad
&\psi \sim \psi' = \psi + \Omega \chi,
\\
&{\rm H}(\,^\star \Omega) = {{\rm Ker} \,^\star\Omega}/
{{\rm Im} \,^\star \Omega},
&\qquad
&\psi \sim \psi' = \psi + \,^\star \Omega \phi.
\end{alignat*}
By imposing only the BRST gauge condition, $\Omega \psi = 0$,
within the equivalence class of BRST--closed states $\psi = \omega +
\Omega \chi$ besides the harmonic state $\omega$ there occur also
spurious BRST--exact states, $\Omega \chi$, which have zero physical
norm. On the other hand, by imposing also the co--BRST gauge condition,
$^\star\Omega \psi = 0$, one gets for each BRST cohomology class the
uniquely determined harmonic state, $\psi = \omega$. Obviously, also the
observables, being functionals of the fields, are elements of the
intersection of the corresponding cohomologies of $\Omega$ and
$^\star \Omega$.

In the following topological gauge theories are called of Hodge--type if their
{\it scalar} supercharges obey a topological superalgebra quite similar to the
BRST complex (\ref{2.1}). More precisely, in these theories (which are
first--stage reducible) the BRST and co--BRST operators are of the form
$\Omega = Q + s$ and $^\star \Omega = \,^\star Q + s$, respectively, where
$Q$ and $^\star Q$ are the generators of the topological
shift and co--shift symmetry, and $s = \delta_G(C)$ is the generator of the
ghost--dependent ordinary gauge transformations ($C$ being the gauge ghosts).
Since it is a common practice to ignore in the gauge fermion
of topological theories that part which fixes the ordinary gauge symmetry,
we always omit in the BRST complex (\ref{2.1}) the part
stemming from the operator $s$. With other words, we look for a topological
superalgebra which has precisely the same form as in (\ref{2.1}), but with
$\Omega$ and $^\star \Omega$ replaced by $Q$ and $^\star Q$, respectively.

\bigskip
\begin{flushleft}
{\large{\bf 3. The topological twist of $N = 2$, $D = 5$
super Yang--Mills theory}}
\end{flushleft}
\medskip

In this section we construct a 5--dimensional cohomological gauge theory
with a simple, $N_T = 1$, scalar supersymmetry $Q$. This topological theory
has the same interesting feature as the Blau--Thompson model
\cite{3} to possess no bosonic scalar fields and hence no underlying
equivariant cohomology. Therefore, it may be considered as a higher
dimensional analogue of that model.

In order to get this theory we first dimensional reduce $N = 1$,
$D = 10$ SYM to $D = 5$ by breaking down the 10--dimensional
Lorentz group according to $SO(1,9) \supset SO(1,4) \otimes
Spin(5)$. The internal symmetry group of this dimensionally
reduced $N = 2$, $D = 5$ SYM is $Spin(5) \sim Sp(4)$, the covering
of the R--symmetry group $SO(5)$. Then, we perform a Wick rotation
to Euclidean space and embed the Euclidean rotation group
$SO_E(5)$ into the global symmetry group such that at least one of
the supercharges of the untwisted theory becomes a scalar with
respect to the new rotation group. This is achieved by taking the
diagonal subgroup of $SO_E(5) \otimes SO(5)$ thereby leading to
the $N_T = 1$ topological twist of the Euclidean $N = 2$, $D = 5$
SYM. Let us stress that it is necessary to start from the
Minkowskian $N = 1$, $D = 10$ SYM because of the well--known fact
that there are no Majorana spinors in Euclidean space. Since the
details of this intrinsically $5$--dimensional twisting procedure
are rather involved we present some of the relevant steps in
detail.

First, the Minkowskian action of $N = 1$, $D = 10$ SYM reads \cite{18},
\begin{equation}
\label{3.1}
S^{(N = 1)} = \int_M d^{10}x\, {\rm tr} \Bigr\{
\hbox{\large$\frac{1}{4}$} F^{MN} F_{MN} -
i \bar{\lambda} \Gamma^M D_M \lambda \Bigr\},
\end{equation}
with $F_{MN} = \partial_M A_N - \partial_N A_M + [ A_M, A_N ]$ and
$D_M = \partial_M + [ A_M, ~\cdot~ ]$. It is build up from an anti--hermitean
vector field $A_M$ ($M = 0, \ldots, 3,5, \ldots, 10$) and a Majorana--Weyl
spinor $\lambda$ in the real $16$--dimensional representation of $SO(1,9)$.
All the fields take their values in the Lie algebra $Lie(G)$ of some compact
gauge group $G$. This action is invariant under the following supersymmetry
transformations (with $16$ real spinorial charges):
\begin{align}
\label{3.2}
&\delta_Q A_M = \bar{\zeta} \Gamma_M \lambda - \bar{\lambda} \Gamma_M \zeta,
\nonumber
\\
&\delta_Q \lambda = \hbox{$\frac{1}{2}$} i \Gamma^M \Gamma^N \zeta F_{MN},
\nonumber
\\
&\delta_Q \bar{\lambda} = - \hbox{$\frac{1}{2}$}
i \bar{\zeta} \Gamma^M \Gamma^N F_{MN},
\end{align}
where $\zeta$ is  a constant Majorana--Weyl spinor. This symmetry is checked
by using the identity
\begin{equation*}
\hbox{$\frac{1}{2}$} \Gamma^{L_0} [ \Gamma^M, \Gamma^N ] =
\eta^{L_0 [M} \Gamma^{N]} - \hbox{$\frac{1}{7!}$}
\epsilon^{L_0 MN L_1 \ldots L_7} \Gamma_{11} \Gamma_{L_1} \ldots \Gamma_{L_7},
\qquad
\Gamma_{11} = \Gamma_0 \ldots \Gamma_{10},
\end{equation*}
where $\eta_{MN} = {\rm diag}(-1,+1, \cdots,+1)$ and
$\epsilon^{L_0 \ldots L_{10}}$ are the metric and the completely antisymmetric
unit tensor in $D = 10$, respectively.

For the $32$--dimensional Dirac matrices, $\{ \Gamma_M, \Gamma_N \} =
2 \eta_{MN} I_{32}$, in  10--dimensional Minkowski space--time we choose the
following block representation:
\begin{align*}
&\Gamma_m = \begin{pmatrix} 0 &
(\gamma_m)_a^{\!~~b} \delta_A^{\!~~B} \\
(\gamma_m)_a^{\!~~b} \delta_A^{\!~~B} & 0 \end{pmatrix},
\qquad\quad~\;
m = 0,1,2,3,5,
\\
&\Gamma_{5 + \alpha} = i \begin{pmatrix} 0 &
\delta_a^{\!~~b} (\gamma_\alpha)_A^{\!~~B} \\
- \delta_a^{\!~~b} (\gamma_\alpha)_A^{\!~~B} & 0 \end{pmatrix},
\qquad
\alpha = 1,2,3,4,5,
\\
&\Gamma_{11} = \begin{pmatrix}
\delta_a^{\!~~b} \delta_A^{\!~~B} & 0 \\ 0 &
- \delta_a^{\!~~b} \delta_A^{\!~~B} \end{pmatrix},
\qquad
C_{10} = \begin{pmatrix} 0 & (C_5)_{ab} \epsilon_{AB} \\
- (C_5)_{ab} \epsilon_{AB} & 0 \end{pmatrix}.
\end{align*}
Here, $(\gamma_m)_a^{\!~~b}$ and $(\gamma_\alpha)_A^{\!~~B}$ are the $SO(1,4)$
and $SO(5)$ matrices, respectively, where both types of spinor indices
($a, b$) and ($A, B$) are taking 4 distinct values,
\begin{align*}
&(\gamma_m)_a^{\!~~b} = ( \gamma_0, \gamma_1, \gamma_2, \gamma_3,
\gamma_5 ),
\qquad
\gamma_5 = i \gamma_0 \gamma_1 \gamma_2 \gamma_3,
\qquad
\gamma_5^2 = I_4,
\\
&(\gamma_\alpha)_A^{\!~~B} = ( \gamma_1, \gamma_2, \gamma_3, \gamma_4,
\gamma_5 ),
\qquad
\gamma_4 = - i \gamma_0.
\end{align*}
$\gamma_\mu$ ($\mu = 0,1,2,3$) and $\gamma_5$ are are chosen to be
equal $-i$ times the usual $SO(1,3)$ Dirac matrices in the Weyl representation
with the metric being $\eta_{\mu\nu} = {\rm diag}(+1,-1,-1,-1)$. Notice,
that the $SO(5)$ matrices $(\gamma_\alpha)_A^{\!~~B}$ are hermitean.

The real, antisymmetric charge conjugation matrix $(C_5)_{ab}$ in the
$5$--dimensional Minkowski space--time has the defining properties
\begin{equation*}
(C_5^{-1})^{ac} (\gamma_m)_c^{\!~~d} (C_5)_{db} =
(\gamma_m)_b^{\!~~a},
\qquad
(C_5)_{ab} = \gamma_5 C_4,
\qquad
C_4 = i \gamma_2 \gamma_0,
\end{equation*}
$C_4$ being the usual $4$--dimensional charge conjugation matrix.
In the $5$--dimensional Euclidean space there exists a real, antisymmetric
tensor $\epsilon_{AB}$, being the invariant tensor of $Sp(4)$, which
transposes the $(\gamma_\alpha)_A^{\!~~B}$ matrices,
\begin{equation*}
\epsilon^{AC} (\gamma_\alpha)_C^{\!~~D} \epsilon_{DB} =
- (\gamma_\alpha)_B^{\!~~A},
\qquad
\epsilon_{AB} = \gamma_3 \gamma_1,
\qquad
\epsilon_{AC} \epsilon^{BC} = \delta_A^{\!~~B},
\end{equation*}
and which can be chosen numerically to be equal to $(C_5)_{ab}$. Moreover,
this tensor can be used as symplectic metric to raise and lower the spinor
index $A$, e.g., $\epsilon^{AC} (\gamma_\alpha)_C^{\!~~B} =
(\gamma_\alpha)^{AB}$ and $(\gamma_\alpha)_A^{\!~~C} \epsilon_{CB} =
(\gamma_\alpha)_{AB}$. According to that convention the matrices
$(\gamma_\alpha)_{AB}$ are antisymmetric,
$(\gamma_\alpha)_{AB} = - (\gamma_\alpha)_{BA}$, and traceless,
$\epsilon^{AB} (\gamma_\alpha)_{AB} = 0$.

The Weyl condition $\lambda = \Gamma_{11} \lambda$ and the
Majorana condition $\lambda = C_{10} \bar{\lambda}^{\rm T}$,
$\bar{\lambda} = \lambda^\dagger \Gamma_0$, restrict a general unconstrained
complex $32$--spinor in the $D = 10$ Minkowski space--time to the real
$16$--spinor $\lambda$. The chirality and the symplectic reality condition
give rise to the structure
\begin{equation*}
\lambda = \begin{pmatrix} \lambda_{a A} \\ 0 \end{pmatrix},
\qquad
\bar{\lambda} = ( 0, \bar{\lambda}^{a A} ),
\qquad
\lambda_{a A} = (C_5)_{ab} \epsilon_{AB} \bar{\lambda}^{b B},
\end{equation*}
i.e., the 16 surviving spinor components $\lambda_{a A}$ are constrained by
a $Sp(4)$--covariant Majorana condition.
We further define
\begin{equation*}
A_M = ( A_m, V_\alpha ),
\end{equation*}
where $V_\alpha$ are the components of the gauge field related to the internal
directions $x^6 \ldots x^{10}$.

As a next step, we compactify 5 of the 10 dimensions by ordinary
dimensional reduction, demanding that no field depends on
$x^6, \ldots, x^{10}$. Then, from (\ref{3.1}) for the dimensionally
reduced action of the Minkowskian $N = 2$, $D = 5$ SYM with
R--symmetry group $SO(5)$ one obtains
\begin{align}
\label{3.3}
S^{(N = 2)} = \int_M d^5x\, {\rm tr} \Bigr\{&
\hbox{$\frac{1}{4}$} F^{mn} F_{mn} +
\hbox{$\frac{1}{2}$} D^m V^\alpha D_m V_\alpha +
\hbox{$\frac{1}{4}$} [ V^\alpha, V^\beta ] [ V_\alpha, V_\beta ]
\nonumber
\\
& - i \bar{\lambda}^{a C} (\gamma^m)_a^{\!~~b} D_m \lambda_{b C} -
\bar{\lambda}^{a C} (\gamma^\alpha)_C^{\!~~D} [ V_\alpha, \lambda_{a D} ]
\Bigr\}.
\end{align}

In order to get from (\ref{3.3}) a cohomological theory we
first perform a Wick rotation, $x_4 = - i x_0$, into the Euclidean space.
Thereby, we relax the symplectic reality conditions and write the
Majorana spinors and their conjugated ones just as in the Minkowskian space,
but consider these spinor fields, from now on, as {\it complex}. Hence,
hermiticity is abandoned.\footnote{Lost of hermiticity in the Euclidean
formulation of a field theory is {\it not} a problem. The primary reason
to impose reality conditions on spinor fields is unitarity, which is
needed only in a field theory with {\it real} time.
We are indebted to the referee for pointing out the lack of reality
of the cohomological action (\ref{3.6}), below.}
Afterwards, we twist the Euclidean rotation group $SO_E(5)$ with
the R--symmetry group $SO(5)$, or, in other words, we identify the spinor
indices $a$ and $A$ (as well as $m$ and $\alpha$). In this way we obtain the
twisted action of the Euclidean $N_T = 1$, $D = 5$ TYM we are looking for,
\begin{align}
\label{3.4}
S^{(N_T = 1)} = \int_E d^5x\, {\rm tr} \Bigr\{&
\hbox{$\frac{1}{4}$} F^{\alpha\beta}(A + i V) F_{\alpha\beta}(A - i V) +
\hbox{$\frac{1}{2}$} D^\alpha(A) V_\alpha D^\beta(A) V_\beta
\nonumber
\\
& - i \lambda^{AC} (\gamma^\alpha)_A^{\!~~B} D_\alpha(A) \lambda_{BC} -
\lambda^{CA} (\gamma^\alpha)_A^{\!~~B} [ V_\alpha, \lambda_{CB} ] \Bigr\},
\end{align}
where now $V_\alpha$ transforms as a co--vector field of $A_\alpha$.

From the transformation rules (\ref{3.2}) one gets the twisted
supersymmetry transformations
\begin{align*}
&\delta_Q A_\alpha = 2 \zeta^{AC} (\gamma_\alpha)_A^{\!~~B} \lambda_{BC},
\\
&\delta_Q V_\alpha = - 2 i \zeta^{CA} (\gamma_\alpha)_A^{\!~~B} \lambda_{CB},
\\
&\delta_Q \lambda_{AB} = - \hbox{$\frac{1}{2}$} i (\sigma_{\alpha\beta})_{AC}
\zeta^C_{\!~~B} F^{\alpha\beta}(A) +
(\gamma_\alpha)_A^{\!~~C} (\gamma_\beta)_B^{\!~~D}
\zeta_{CD} D^\alpha(A) V^\beta -
\hbox{$\frac{1}{2}$} i (\sigma_{\alpha\beta})_{BC} \zeta_A^{\!~~C}
[ V^\alpha, V^\beta ].
\end{align*}
Since there occur no half--integer spin fields in the action (\ref{3.4}) we
can convert the spinor notation into the more familar tensor notation by
decomposing the twisted spinor fields $\lambda_{AB}$ as follows,
\begin{equation}
\label{3.5}
\lambda_{AB} =  \hbox{$\frac{1}{2\sqrt{2}}$} \Bigr\{
(\gamma^\alpha)_{AB} \psi_\alpha +
\hbox{$\frac{1}{2}$} (\sigma^{\alpha\beta})_{AB} \chi_{\alpha\beta} -
\epsilon_{AB} \tilde{\eta} \Bigr\},
\qquad
\lambda^{AB} = \epsilon^{AC} \epsilon^{BD} \lambda_{CD},
\end{equation}
where $\tilde{\eta}$, $\psi_\alpha$ and $\chi_{\alpha\beta}$ are
Grassmann--odd ghost--for--antighost scalar, vector and antisymmetric
(traceless) tensor fields, respectively. Here, the 10 generators
$(\sigma_{\alpha\beta})_{AB}$ of the $Sp(4)$ rotations obey the relations
\begin{align*}
&(\gamma_\alpha)_A^{\!~~C} (\gamma_\beta)_{CB} =
\delta_{\alpha\beta} \epsilon_{AB} - (\sigma_{\alpha\beta})_{AB},
\\
&(\gamma_\alpha)_A^{~~\!C} (\sigma_{\beta\gamma})_{CB} =
\delta_{\alpha\gamma} (\gamma_\beta)_{AB} -
\delta_{\alpha\beta} (\gamma_\gamma)_{AB} -
\hbox{$\frac{1}{2}$} \epsilon_{\alpha\beta\gamma\delta\eta}
(\sigma^{\delta\eta})_{AB},
\\
&(\sigma_{\alpha\beta})_A^{\!~~C} (\sigma_{\gamma\delta})_{CB} =
\delta_{\alpha\gamma} (\sigma_{\beta\delta})_{AB} -
\delta_{\alpha\delta} (\sigma_{\beta\gamma})_{AB} +
\delta_{\beta\delta} (\sigma_{\alpha\gamma})_{AB} -
\delta_{\beta\gamma} (\sigma_{\alpha\delta})_{AB}
\\
&\phantom{(\sigma_{\alpha\beta})_A^{\!~~C} (\sigma_{\gamma\delta})_{CB} =}
+ ( \delta_{\alpha\delta} \delta_{\beta\gamma} -
\delta_{\alpha\gamma} \delta_{\beta\delta} ) \epsilon_{AB} -
\epsilon_{\alpha\beta\gamma\delta\eta} (\gamma^\eta)_{AB},
\end{align*}
where $\epsilon_{\alpha\beta\gamma\delta\eta}$ is the completely
antisymmetric unit tensor in $D = 5$. Notice, that from the first, defining
relation it follows that $(\sigma_{\alpha\beta})_{AB}$, owing to the
antisymmetry of $(\gamma_\alpha)_{AB}$, is symmetric,
$(\sigma_{\alpha\beta})_{AB} = (\sigma_{\alpha\beta})_{BA}$.

Inserting the decomposition (\ref{3.5}) into (\ref{3.4}) and introducing
the bosonic auxiliary field $Y$, the virtue of which is to make the
topological supersymmetry $Q$ strictly nilpotent, for the resulting
action, with an underlying non--equivariant $N_T = 1$ topological
supersymmetry $Q$, we get
\begin{align}
\label{3.6}
S^{(N_T = 1)} = \int_E d^5x\, {\rm tr} \Bigr\{&
\hbox{$\frac{1}{4}$} F^{\alpha\beta}(A + i V) F_{\alpha\beta}(A - i V) -
\hbox{$\frac{1}{4}$} i \tilde{\chi}^{\alpha\beta\gamma}
D_\alpha(A - i V) \chi_{\beta\gamma}
\nonumber
\\
& - i \chi^{\alpha\beta} D_\alpha(A + i V) \psi_\beta -
i \psi^\alpha D_\alpha(A - i V) \tilde{\eta} -
Y D^\alpha(A) V_\alpha - \hbox{$\frac{1}{2}$} Y^2 \Bigr\};
\end{align}
$\tilde{\chi}_{\alpha\beta\gamma} = \frac{1}{2}
\epsilon_{\alpha\beta\gamma\delta\eta} \chi^{\delta\eta}$ is the dual
of $\chi_{\alpha\beta}$. This action can be written as a sum of a topological
term ($Q$--cocycle) and a $Q$--exact term,
\begin{equation}
\label{3.7}
S^{(N_T = 1)} = - \int_E d^5x\, {\rm tr} \Bigr\{
\hbox{$\frac{1}{4}$} i \tilde{\chi}^{\alpha\beta\gamma}
D_\alpha(A - i V) \chi_{\beta\gamma} \Bigr\} + Q \Psi,
\qquad
Q^2 = 0,
\end{equation}
with the gauge fermion
\begin{equation*}
\Psi = \int_E d^5x\, {\rm tr} \Bigr\{
\hbox{$\frac{1}{4}$} i \chi^{\alpha\beta} F_{\alpha\beta}(A + i V) -
\tilde{\eta} D^\alpha(A) V_\alpha - \hbox{$\frac{1}{2}$} \tilde{\eta} Y
\Bigr\}.
\end{equation*}
The topological supersymmetry transformations, generated by $Q$, are given by
\begin{alignat*}{2}
&Q A_\alpha = \psi_\alpha,
&\qquad\qquad
&Q V_\alpha = - i \psi_\alpha,
\\
&Q \psi_\alpha = 0,
&\qquad\qquad
&Q \chi_{\alpha\beta} = - i F_{\alpha\beta}(A - i V),
\\
&Q \tilde{\eta} = Y,
&\qquad\qquad
&Q Y = 0.
\end{alignat*}
This $N_T = 1$ topological model in $D = 5$ bears a strong resemblance to
the novel $N_T = 2$ topological twist of $N = 4$, $D = 3$ SYM constructed by
Blau and Thompson \cite{3}. The most striking feature of this model is that
the topological supersymmetry $Q$ is not equivariantly nilpotent but rather
strictly nilpotent, i.e., prior to the introduction of the gauge ghosts $C$.
Moreover, this model, as promised, has no bosonic scalar fields. Furthermore,
the vector field $V_\alpha$ is combined with $A_\alpha$ to form the
complexified gauge field $A_\alpha \pm i V_\alpha$, where
$A_\alpha - i V_\alpha$ is $Q$--invariant, which is another remarkable
property of that theory.

Since $Q$ is a $Sp(4)$--singlet the twisted action (\ref{3.6}) can be put onto
a general $5$--dimensional Riemannian manifold with Euclidean signature and
$SO(5)$ holonomy group. In this way one gets a topological theory in $D = 5$
whose key property is the $Q$--exactness of its stress tensor, which implies
that the correlation functions of $Q$--invariant observables are independent
of the metric of the manifold.

We conjecture that this higher dimensional analogue of the Blau--Thompson
model is the only topological one which can be constructed in $D = 5$.
In the next Section it will be shown that from (\ref{3.6}), after dimensional
reduction to $D = 4$, one obtains precisely the action of the B--model
\cite{11,3}, a certain topological twist of $N = 4$ SYM, which
localizes onto the moduli space of complexified flat connections. However, it
was found that there are two more topological twists of $N = 4$ SYM,
namely the A--model \cite{19,20,3} and the half--twisted model \cite{20},
whose actions localize onto the moduli space of instantons.
Thus, one should expect that in $D = 5$, apart from the topological model
constructed above, there are at least two more cohomological gauge theories
as well. But, both these theories are neither topological nor are they twisted
versions of $N = 2$, $D = 5$ SYM. Rather they are untwisted $N = 2$ SYM
theories formulated on manifolds with reduced holonomy group
$H \subset SO(5)$. Namely, since instantons in $D = 5$ require the existence
of a Hodge self--dual 4--form it must be a singlet with respect to a proper
subgroup of $SO(5)$ and not of $SO(5)$ itself. Therefore, these theories are
only invariant under a certain class of metric variations which preserve the
(reduced) holonomy. Perhaps, the simplest way to obtain both these theories
is to perform a dimensional reduction to $D = 5$ of the higher dimensional
analogue of the Donaldson--Witten theory in $D = 8$ and $D = 7$ on a Joyce
manifold with $Spin(7)$ and $G_2$ holonomy \cite{8,9}, respectively.

Moreover, it would be interesting to study the relationship between the
higher dimensional analogue of the Blau--Thompson model and the dimensionally
reduced $N_T = 2$ and $N_T = 3$ cohomological gauge theories on a Calabi--Yau
4--fold and on a quartionic K\"ahler manifold with $SU(4)$ and
$Sp(4) \otimes Sp(2)$ holonomy \cite{8,10}, respectively, to $D = 5$.
However, here we will not further dwell on these issues.

\bigskip
\begin{flushleft}
{\large{\bf 4. The $N_T = 2$, $D = 4$ topological B--model and the
$N_T = 4$, $D = 3$ equivariant extension of the Blau--Thompson model}}
\end{flushleft}
\medskip

We are now going to discuss the relation between the $N_T = 1$, $D = 5$
topological model constructed above and the $N_T = 4$ equivariant
extension \cite{12} of the Blau--Thompson model in $D = 3$. To this end we
first perform a $(4 + 1)$--decomposition of the action (\ref{3.6}), i.e.,
we split the coordinates into $x^\alpha = ( x^\mu, x^5 )$, $\mu = 1,2,3,4$.
Furthermore, we assume that no field depends on $x^5$, i.e., $\partial_5 = 0$.
As a next step, we rename the fifth component of $A_\alpha \pm i V_\alpha$,
$\chi_{\alpha\beta}$ and $\psi_\alpha$ according to
\begin{equation*}
A_5 - i V_5 = G,
\qquad
A_5 + i V_5 = \bar{G},
\qquad
\chi_{\mu 5} = \tilde{\psi}_\mu,
\qquad
\psi_5 = \eta,
\end{equation*}
reserving the notation $A_\mu \pm i V_\mu$, $\chi_{\mu\nu}$ and $\psi_\mu$
for the corresponding fields in $D = 4$. Then, after squeezing (\ref{3.6})
to $D = 4$, we arrive precisely at the action of the topological B--model,
with an extended $N_T = 2$ on--shell equivariantly nilpotent topological
supersymmetry, constructed by Marcus \cite{11},
\begin{align}
\label{4.1}
S^{(N_T = 2)} = \int_E d^4x\, {\rm tr} \Bigr\{&
\hbox{$\frac{1}{4}$} F^{\mu\nu}(A + i V) F_{\mu\nu}(A - i V) +
\hbox{$\frac{1}{4}$} D^\mu(A + i V) G D_\mu(A - i V) \bar{G}
\nonumber
\\
& + \hbox{$\frac{1}{4}$} D^\mu(A - i V) G D_\mu(A + i V) \bar{G} -
\hbox{$\frac{1}{8}$} [ G, \bar{G} ]^2 -
Y D^\mu(A) V_\mu - \hbox{$\frac{1}{2}$} Y^2
\nonumber
\\
& - i \chi^{\mu\nu} D_\mu(A + i V) \psi_\nu -
i \tilde{\chi}^{\mu\nu} D_\mu(A - i V) \tilde{\psi}_\nu +
\hbox{$\frac{1}{4}$} i G \{ \chi^{\mu\nu}, \tilde{\chi}_{\mu\nu} \}
\phantom{\frac{1}{2}}
\nonumber
\\
& - i \tilde{\psi}^\mu D_\mu(A + i V) \eta -
i \psi^\mu D_\mu(A - i V) \tilde{\eta} -
i \bar{G} \{ \psi^\mu, \tilde{\psi}_\mu \} +
i G \{ \eta, \tilde{\eta} \} \Bigr\},
\end{align}
where $\tilde{\chi}_{\mu\nu} = \hbox{$\frac{1}{2}$}
\epsilon_{\mu\nu\rho\sigma} \chi^{\rho\sigma}$ is the dual of $\chi_{\mu\nu}$.
This topological action, which localizes onto the moduli space of the
complexified gauge fields $A_\mu \pm i V_\mu$, can be regarded also
as a deformation of the $N_T = 1$, $D = 4$ super--BF theory \cite{3}.
Notice that in (\ref{4.1}) hermiticity is restored.

Obviously, the action (\ref{4.1}) is invariant under the following
discrete $Z_2$ symmetry,
\begin{equation*}
Z_2:
\qquad
\begin{bmatrix}
A_\mu & V_\mu & G & \bar{G} & Y & \\
\chi_{\mu\nu} & \tilde{\chi}_{\mu\nu} & \psi_\mu & \tilde{\psi}_\mu &
\eta & \tilde{\eta} \end{bmatrix}
\quad\Rightarrow\quad
\begin{bmatrix}
A_\mu & - V_\mu & G & \bar{G} & - Y & \\
\tilde{\chi}_{\mu\nu} & \chi_{\mu\nu} & \tilde{\psi}_\mu & \psi_\mu &
\tilde{\eta} & \eta \end{bmatrix},
\end{equation*}
which also maps the topological supercharge $Q$ into $\tilde{Q}$, i.e., the
topological supersymmetry is actually $N_T = 2$.
\smallskip

Let us now perform a further dimensional reduction of this topological
B--model to $D = 3$. For that purpose we introduce a $SU(2)$-- and a
$\overline{SU(2)}$--doublet of Grassmann--odd ghost--for--antighost scalar
and vector fields, $\eta^a$, $\psi_\alpha^a$ and
$\bar{\eta}^a$, $\bar{\psi}_\alpha^a$, respectively, and a
$SU(2) \otimes \overline{SU(2)}$--quartet of Grassmann--even complex scalar
fields $M^{ab}$, according to
\begin{align}
\label{4.2}
&\psi_\alpha^a = \begin{pmatrix} \tilde{\chi}_{4 \alpha} \\
\psi_\alpha \end{pmatrix},
\qquad
\bar{\psi}_\alpha^a = i \begin{pmatrix} \chi_{4 \alpha} \\
\tilde{\psi}_\alpha \end{pmatrix},
\qquad
\eta^a = \begin{pmatrix} \eta \\ \psi_4 \end{pmatrix},
\qquad
\bar{\eta}^a = i \begin{pmatrix} \tilde{\eta} \\
\tilde{\psi}_4 \end{pmatrix},
\nonumber
\\
&M^{ab} = i \begin{pmatrix} \bar{G} & A_4 - i V_4 \\
A_4 + i V_4 & - G \end{pmatrix},
\qquad
G = A_5 - i V_5,
\qquad
\bar{G} = A_5 + i V_5,
\end{align}
where the space index $\alpha$ runs now between 1 and 3. The internal
group index $a = 1,2$ is raised and lowered as follows:
$\epsilon^{ac} \varphi_c^{~b} = \varphi^{ab}$ and
$\varphi_a^{~c} \epsilon_{cb} = \varphi_{ab}$, where $\epsilon^{ab}$ is the
invariant tensor of the group $SU(2)$, $\epsilon_{12} = \epsilon^{12} = 1$.
The matrices $M^{ab}$ and $M_{ab}$ can be rewritten
as follows
\begin{equation*}
M^{ab} = (\sigma_m)^{ab} M^m,
\qquad
M_{ab} = (\sigma_m^*)^{ab} M^m =
M^{cd} \epsilon_{ca} \epsilon_{db},
\qquad
m = 1,2,3,4,
\end{equation*}
with $M^m = \{ A_{4,5}, V_{4,5} \}$, where $(\sigma_m)^{ab} =
( i \sigma_1, i \sigma_2, i \sigma_3, - I_2 )$ and its hermitean conjugate
are the Clebsch--Gordon coefficients relating the vector representation of
$SO(4)$ to the $(1/2,1/2)$ representation of $SU(2) \otimes \overline{SU(2)}$,
$\sigma_\alpha$ ($\alpha = 1,2,3$) being the Pauli matrices. Moreover,
in order to close the topological superalgebra off--shell, we introduce an
additional set of bosonic auxiliary fields, namely the complex vector fields
$B_\alpha$ and $\bar{B}_\alpha$.

As a result, after that dimensional reduction from the action (\ref{4.1})
one gets the $N_T = 4$, $D = 3$ off--shell equivariant extension of the
Blau--Thompson model with global symmetry group
$SU(2) \otimes \overline{SU(2)}$,
\begin{align}
\label{4.3}
S^{(N_T = 4)} = \int_E d^3x\, {\rm tr} \Bigr\{&
\hbox{$\frac{1}{4}$} i \epsilon^{\alpha\beta\gamma}
B_\gamma F_{\alpha\beta}(A + i V) -
\hbox{$\frac{1}{4}$} i \epsilon^{\alpha\beta\gamma}
\bar{B}_\gamma F_{\alpha\beta}(A - i V) -
\hbox{$\frac{1}{2}$} \bar{B}^\alpha B_\alpha
\nonumber
\\
& + \hbox{$\frac{1}{2}$} i \epsilon^{\alpha\beta\gamma}
\epsilon_{ab} \psi_\gamma^a D_\alpha(A + i V) \psi_\beta^b -
\epsilon_{ab} \eta^a D^\alpha(A + i V) \bar{\psi}_\alpha^b
\nonumber
\\
& + \hbox{$\frac{1}{4}$} D^\alpha(A + i V) M_{ab} D_\alpha(A - i V) M^{ab} -
Y D^\alpha(A) V_\alpha - \hbox{\large$\frac{1}{2}$} Y^2
\phantom{\frac{1}{2}}
\nonumber
\\
& - \hbox{$\frac{1}{2}$} i \epsilon^{\alpha\beta\gamma}
\epsilon_{ab} \bar{\psi}_\gamma^a D_\alpha(A - i V) \bar{\psi}_\beta^b  -
\epsilon_{ab} \bar{\eta}^a D^\alpha(A - i V) \psi_\alpha^b
\nonumber
\\
& + \hbox{$\frac{1}{16}$} [ M_{ab}, M_{cd} ] [ M^{ab}, M^{cd} ] +
i M_{ab} \{ \bar{\psi}^{\alpha a}, \psi_\alpha^b \} +
i M_{ab} \{ \eta^a, \bar{\eta}^b \} \Bigr\}.
\end{align}
The $N_T = 4$ topological supercharges will be denoted by $Q^a$ and
$\bar{Q}^a$. They are interchanged by the following discrete $Z_2$ symmetry
of the action (\ref{4.3}),
\begin{align*}
Z_2:
\qquad
\begin{bmatrix}
A_\alpha & V_\alpha & B_\alpha & \bar{B}_\alpha & Y \\
\psi_\alpha^a & \bar{\psi}_\alpha^a & \eta^a & \bar{\eta}^a &
M^{ab} \end{bmatrix}
\quad\Rightarrow\quad
\begin{bmatrix}
A_\alpha & - V_\alpha & - \bar{B}_\alpha & - B_\alpha & - Y \\
i \bar{\psi}_\alpha^a & - i \psi_\alpha^a & i \bar{\eta}^a & - i \eta^a &
M^{ba} \end{bmatrix}.
\end{align*}
Thereby, $Q^a$ and $\bar{Q}^a$ transform as a doublet of $SU(2)$ and
$\overline{SU(2)}$, respectively. The transformation laws of the topological
supersymmetry $Q^a$ are
\begin{alignat}{2}
\label{4.4}
&Q^a A_\alpha = \psi_\alpha^a,
&\qquad\qquad
&Q^a V_\alpha = - i \psi_\alpha^a,
\nonumber
\\
&Q^a \psi_\alpha^b = \epsilon^{ab} B_\alpha,
&\qquad \qquad
&Q^a \bar{\psi}_\alpha^b = i D_\alpha(A - i V) M^{ba},
\nonumber
\\
&Q^a \eta^b = 0,
&\qquad\qquad
&Q^a \bar{\eta}^b = - i \epsilon^{ab} Y +
\hbox{$\frac{1}{2}$} \epsilon_{cd} [ M^{ca}, M^{db} ],
\nonumber
\\
&Q^a M^{bc} = - 2 i \epsilon^{ac} \eta^b,
&\qquad\qquad
&Q^a Y = - \epsilon_{cd} [ M^{ca}, \eta^d ],
\nonumber
\\
&Q^a B_\alpha = 0,
&\qquad\qquad
&Q^a \bar{B}_\alpha = - 2 D_\alpha(A - i V) \bar{\eta}^a -
2 i \epsilon_{cd} [ M^{ca}, \bar{\psi}_\alpha^d ].
\end{alignat}
Combining (\ref{4.4}) with the $Z_2$--symmetry, which maps $Q^a$ into
$\bar{Q}^a$, one gets the corresponding transformation laws of $\bar{Q}^a$.
It is straightforward to prove that indeed
\begin{equation*}
( Q^a, \bar{Q}^a ) S^{(N_T = 4)} = 0.
\end{equation*}
Furthermore, by an explicit calculation it can be verified that
$Q^a$ and $\bar{Q}^a$ are both strictly nilpotent and anticommute with each
other modulo the field--dependent gauge transformation $\delta_G(M^{ab})$,
i.e., they satisfy the topological superalgebra off--shell,
\begin{equation*}
\{ Q^a, Q^b \} = 0,
\qquad
\{ \bar{Q}^a, Q^b \} = - 2 \delta_G(M^{ab}),
\qquad
\{ \bar{Q}^a, \bar{Q}^b \} = 0~;
\end{equation*}
here, the gauge transformations are defined by $\delta_G(\varphi) A_\alpha =
- D_\alpha \varphi$ and $\delta_G(\varphi) X = [ \varphi, X ]$ for all the
other fields. This algebraic structure looks like the BRST complex (\ref{2.1}).
However, both operators, $Q^a$ and  $\bar Q^a$, are interrelated by the $Z_2$
symmetry and not, as it should be, by any Hodge--type $\star$ operation.

Finally, let us mention that precisely the same topological action (\ref{4.3})
arises from a dimensional reduction of the A--model \cite{16}
to $D = 3$ \cite{3,12}.

\bigskip
\begin{flushleft}
{\large{\bf 5. $N_T = 8$, $D = 2$ Hodge--type cohomological gauge theory
with global symmetry group $SU(2) \otimes \overline{SU(2)}$}}
\end{flushleft}
\medskip

Now, we come to the main objective of that paper, namely we show that by a
further dimensional reduction of the action (\ref{4.3}) to $D = 2$ we obtain
another example of a Hodge--type cohomological gauge theory whose $N_T = 8$
first--stage reducible gauge symmetries are fixed by harmonic gauges. Indeed,
the $N_T = 4$ topological shift symmetries $Q^a$, $\bar{Q}^a$ and the
$N_T = 4$ topological co--shift symmetries $^\star Q^a = - P \star Q^a \star$,
$^\star \bar{Q}^a = - P \star \bar{Q}^a \star$, being interchanged by a
discrete Hodge--type $\star$ operation, obey the BRST complex (\ref{2.1})
(see, Eq. (\ref{5.4}) below). Here, and in the following $P$ denotes the
operator of Grassmann--parity whose both eigenvalues $\pm 1$ are defined
through
\begin{equation*}
P \varphi = \begin{cases} + \varphi & \text{if $\varphi$ is Grassmann--even},
            \\            - \varphi & \text{if $\varphi$ is Grassmann--odd}.
            \end{cases}
\end{equation*}
To begin with, we rename the third component of $A_\alpha \pm i V_\alpha$,
$\psi_\alpha^a$ and $\bar{\psi}_\alpha^a$ according to
\begin{equation}
\label{5.1}
A_3 - i V_3 = M,
\qquad
A_3 + i V_3 = \bar{M},
\qquad
\psi_3^a = i \bar{\zeta}^a,
\qquad
\bar{\psi}_3^a = - i \zeta^a.
\end{equation}
In order to close the topological superalgebra off--shell we introduce an
additional set of bosonic auxiliary fields, namely the complex vector fields
$E_\mu$, $\bar{E}_\mu$ and the $SU(2) \otimes \overline{SU(2)}$--quartet of
complex vector fields $E_\mu^{ab}$, where now $\mu$ denotes the space
index taking the values 1 and 2.

Then, after performing in the action (\ref{4.3}) the dimensional reduction to
$D = 2$ we arrive at the following $N_T = 8$ Hodge--type cohomological gauge
theory with global symmetry group $SU(2) \otimes \overline{SU(2)}$,
\begin{align}
\label{5.2}
S^{(N_T = 8)} = \int_E & d^2x\, {\rm tr} \Bigr\{
\hbox{$\frac{1}{4}$} i \epsilon^{\mu\nu} B F_{\mu\nu}(A + i V) -
\hbox{$\frac{1}{4}$} i \epsilon^{\mu\nu} \bar{B} F_{\mu\nu}(A - i V) -
\hbox{$\frac{1}{2}$} \bar{B} B
\nonumber
\\
& - \epsilon^{\mu\nu} \epsilon_{ab} \bar{\zeta}^a
D_\mu(A + i V) \psi_\nu^b +
\hbox{$\frac{1}{2}$} i \epsilon^{\mu\nu} \epsilon_{ab} \bar{M}
\{ \psi_\mu^a, \psi_\nu^b \} +
i \epsilon_{ab} M \{ \bar{\eta}^a, \bar{\zeta}^b \}
\nonumber
\\
& - \epsilon_{ab} \bar{\eta}^a D^\mu(A - i V) \psi_\mu^b +
\hbox{$\frac{1}{4}$} D^\mu(A - i V) M D_\mu(A + i V) \bar{M}
\nonumber
\phantom{\frac{1}{2}}
\\
& + \hbox{$\frac{1}{4}$} D^\mu(A + i V) M_{ab} D_\mu(A - i V) M^{ab} -
Y D^\mu(A) V_\mu - \hbox{$\frac{1}{2}$} Y^2
\nonumber
\\
& - \epsilon^{\mu\nu} \epsilon_{ab} \zeta^a
D_\mu(A - i V) \bar{\psi}_\nu^b -
\hbox{$\frac{1}{2}$} i \epsilon^{\mu\nu} \epsilon_{ab} M
\{ \bar{\psi}_\mu^a, \bar{\psi}_\nu^b \} -
i \epsilon_{ab} \bar{M} \{ \eta^a, \zeta^b \}
\nonumber
\phantom{\frac{1}{2}}
\\
& - \epsilon_{ab} \eta^a D^\mu(A + i V) \bar{\psi}_\mu^b +
\hbox{$\frac{1}{4}$} D^\mu(A + i V) M D_\mu(A - i V) \bar{M}
\nonumber
\\
& + \hbox{$\frac{1}{16}$} [ M_{ab}, M_{cd} ] [ M^{ab}, M^{cd} ] +
\hbox{$\frac{1}{4}$} [ M, M_{ab} ] [ \bar{M}, M^{ab} ] -
\hbox{$\frac{1}{8}$} [ M, \bar{M} ]^2
\nonumber
\phantom{\frac{1}{2}}
\\
& + i M_{ab} \{ \bar{\psi}^{\mu a}, \psi_\mu^b \} +
i M_{ab} \{ \eta^a, \bar{\eta}^b \} +
i M_{ab} \{ \zeta^a, \bar{\zeta}^b \} -
\bar{E}^\mu E_\mu - \hbox{$\frac{1}{2}$} E^\mu_{ab} E_\mu^{ab} \Bigr\}.
\end{align}
Let us recall that in the $D = 2$ dimensional Euclidean space there are no
propagating degrees of freedom associated with $A_\mu$.

The action (\ref{5.2}) is manifestly invariant under the following discrete
$Z_2$ symmetry,
\begin{equation*}
Z_2:
\qquad
\begin{bmatrix}
A_\mu & V_\mu & &
\\
B & \bar{B} & Y &
\\
\psi_\mu^a & \eta^a & \zeta^a & M^{ab}
\\
\bar{\psi}_\mu^a & \bar{\eta}^a & \bar{\zeta}^a & E_\mu^{ab}
\\
M & \bar{M} & E_\mu & \bar{E}_\mu
\end{bmatrix}
\quad \Rightarrow \quad
\begin{bmatrix} A_\mu & - V_\mu & &
\\
- \bar{B} & - B & - Y &
\\
i \bar{\psi}_\mu^a & i \bar{\eta}^a & i \bar{\zeta}^a & M^{ba}
\\
- i \psi_\mu^a & - i \eta^a & - i \zeta^a & - E_\mu^{ba}
\\
\bar{M} & M & - \bar{E}_\mu & - E_\mu
\end{bmatrix},
\end{equation*}
mapping $Q^a$ into $\bar{Q}^a$ as well as $^\star Q^a$ into
$^\star \bar{Q}^a$, which will be defined immediatly.

In addition, the action (\ref{5.2}) is also a invariant under the following
discrete Hodge--type $\star$ symmetry, defined by the replacements
\begin{equation*}
\varphi \equiv \begin{bmatrix}
\partial_\mu & A_\mu & V_\mu &
\\
B & \bar{B} & Y &
\\
\psi_\mu^a & \eta^a & \zeta^a & M^{ab}
\\
\bar{\psi}_\mu^a & \bar{\eta}^a & \bar{\zeta}^a & E_\mu^{ab}
\\
M & \bar{M} & E_\mu & \bar{E}_\mu
\end{bmatrix}
\quad \Rightarrow \quad
\star \varphi = \begin{bmatrix} \epsilon_{\mu\nu} \partial^\nu &
\epsilon_{\mu\nu} A^\nu & - \epsilon_{\mu\nu} V^\nu &
\\
- \bar{B} & - B & - Y &
\\
- i \psi_\mu^a & - i \zeta^a & i \eta^a & - M^{ab}
\\
- i \bar{\psi}_\mu^a & - i \bar{\zeta}^a & i \bar{\eta}^a &
\epsilon_{\mu\nu} E^{\nu ab}
\\
- M & - \bar{M} & \epsilon_{\mu\nu} E^\nu & \epsilon_{\mu\nu} \bar{E}^\nu
\end{bmatrix},
\end{equation*}
with the property $\star (\star \varphi) = P \varphi$, mapping $Q^a$ and
$\bar{Q}^a$ into $^\star Q^a = - P \star Q^a \star$ and
$^\star \bar{Q}^a = - P \star \bar{Q}^a \star$, respectively.

The topological shift transformations, generated by $Q^a$, are given as
follows
\begin{align}
\label{5.3}
&Q^a A_\mu = \psi_\mu^a,
\nonumber
\\
&Q^a V_\mu = - i \psi_\mu^a,
\nonumber
\\
&Q^a M = 0,
\nonumber
\\
&Q^a \bar{\zeta}^b = - i \epsilon^{ab} B,
\nonumber
\\
&Q^a B = 0,
\nonumber
\\
&Q^a M^{bc} = - 2 i \epsilon^{ac} \eta^b,
\nonumber
\\
&Q^a \eta^b = 0,
\nonumber
\\
\intertext{}
&Q^a \psi_\mu^b = \epsilon^{ab} E_\mu -
i \epsilon^{ab} \epsilon_{\mu\nu} D^\nu(A - i V) M,
\nonumber
\\
&Q^a E_\mu = 0,
\nonumber
\\
&Q^a \bar{\psi}_\mu^b = \epsilon_{\mu\nu} E^{\nu ba} +
i D_\mu(A - i V) M^{ba},
\nonumber
\\
&Q^a \bar{M} = 2 i \bar{\zeta}^a,
\nonumber
\\
&Q^a \bar{\eta}^b = - i \epsilon^{ab} Y +
\hbox{$\frac{1}{2}$} \epsilon^{ab} [ M, \bar{M} ] +
\hbox{$\frac{1}{2}$} \epsilon_{cd} [ M^{ca}, M^{db} ],
\nonumber
\\
&Q^a \zeta^b = [ M^{ba}, M ],
\nonumber
\\
&Q^a Y = [ M, \bar{\zeta}^a ] - \epsilon_{cd} [ M^{ca}, \eta^d ],
\nonumber
\\
&Q^a \bar{E}_\mu = \epsilon_{\mu\nu} D^\nu(A + i V) \bar{\zeta}^a +
i \epsilon_{\mu\nu} [ \bar{M}, \psi^{\nu a} ] -
D_\mu(A - i V) \bar{\eta}^a - i \epsilon_{cd} [ M^{ca}, \bar{\psi}_\mu^d ],
\nonumber
\\
&Q^a \bar{B} = 2 [ \bar{\eta}^a, M ] - 2 \epsilon_{cd} [ M^{ca}, \zeta^d ],
\nonumber
\\
&Q^a E_\mu^{bc} = - \epsilon^{ac} \epsilon_{\mu\nu} D^\nu(A + i V) \eta^b -
i \epsilon^{ac} \epsilon_{\mu\nu} [ M^b_{~d}, \psi^{\nu d} ] -
\epsilon^{ac} D_\mu(A - i V) \zeta^b +
i \epsilon^{ac} [ M, \bar{\psi}_\mu^b ].
\end{align}
Combining these transformation rules (\ref{5.3}) with the $Z_2$ and/or the
Hodge--type $\star$ symmetry one obtains the complete set of symmetry
transformations which leave the action (\ref{5.2}) invariant. Indeed, by a
rather lengthly calculation one verifies that
\begin{equation*}
( Q^a, \bar{Q}^a, \,^\star Q^a, \,^\star \bar{Q}^a ) S^{(N_T = 8)} = 0.
\end{equation*}
Furthermore, after a straightforward, but tedious calculation one also
verifies that $Q^a$, $\bar{Q}^a$ and $^\star Q^a$, $^\star \bar{Q}^a$
provide an off--shell realization of the following topological superalgebra,
\begin{alignat}{3}
\label{5.4}
&\{ Q^a, Q^b \} = 0,
&\qquad
&\{ Q^a, \,^\star Q^b \} = - 2 \epsilon^{ab} \delta_G(M),
&\qquad
&\{ \,^\star Q^a, \,^\star Q^b \} = 0,
\nonumber
\\
&\{ Q^a, \,^\star \bar{Q}^b \} = 0,
&\qquad
&\{ Q^a, \bar{Q}^b \} = - 2 \delta_G(M^{ba}),
&\qquad
&\{ \,^\star Q^a, \bar{Q}^b \} = 0,
\nonumber
\\
&\{ \bar{Q}^a, \,^\star Q^b \} = 0,
&\qquad
&\{ \,^\star \bar{Q}^a, \,^\star Q^b \} = 2 \delta_G(M^{ab}),
&\qquad
&\{ \,^\star \bar{Q}^a, Q^b \} = 0,
\nonumber
\\
&\{ \bar{Q}^a, \bar{Q}^b \} = 0,
&\qquad
&\{ \,^\star \bar{Q}^a, \bar{Q}^b \} = 2 \epsilon^{ab} \delta_G(\bar{M}),
&\qquad
&\{ \,^\star \bar{Q}^a, \,^\star \bar{Q}^b \} = 0.
\end{alignat}
Obviously, this superalgebra is analogous to the de Rham cohomology in
differential geometry: The nilpotent
topological shift and co--shift operators, $( Q^a, \bar{Q}^a )$ and
$( \,^\star Q^a, \,^\star \bar{Q}^a )$, correspond to the exterior and the
co--exterior derivatives, $d$ and $\delta = \pm \star d \star$, respectively,
where both are interrelated by the Hodge--type $\star$ operations.
Furthermore, the various field--dependent gauge generators
$\delta_G( M, \bar{M}, M^{ab} )$, correspond to the Laplacian
$\Delta = \{ d, \delta \}$ so that we have indeed a perfect example of a
Hodge--type cohomological gauge theory in $D = 2$.

\bigskip
\begin{flushleft}
{\large{\bf 6. $N_T = 8$, $D = 2$ Hodge--type cohomological gauge theory
with global symmetry group $SU(4)$}}
\end{flushleft}
\medskip

The $Z_2$ symmetry of the action (\ref{5.2}) is immediately related to the
factorization of the global symmetry group into both the subgroups $SU(2)$
and $\overline{SU(2)}$. Now, we want to show that this action can be cast
into a form where it actually has the larger global symmetry group $SU(4)$.
This is in accordance with the group theoretical description of some classes
of topologically twisted low--dimensional supersymmetric world--volume
theories given in \cite{3}. Such effective low--energy world--volume
theories appear quite naturally in the study of curved D--branes and
D--brane instantons wrapping around supersymmetric cocycles for special
Lagrangian submanifolds of Calabi--Yau $n$--folds \cite{13,14,3}.

Indeed, the structure of the superalgebra (\ref{5.4}) suggests that the
6 scalar fields ($M$, $\bar{M}$, $M^{ab}$) could be combined to form a sextet
of the group $SO(6) \sim SU(4)$ which should be a global symmetry group of
the theory. To elaborate this suggestion we introduce a $SU(4)$--quartet of
Grassmann--odd vector fields $\psi_\mu^\alpha$, two $SU(4)$--quartets of
Grassmann--odd scalar fields, $\bar{\eta}_\alpha$ and $\bar{\zeta}_\alpha$,
which transform as the fundamental and its complex conjugate representation
of $SU(4)$, respectively, and a $SU(4)$--sextet of Grassmann--even complex
scalar fields $M_{\alpha\beta} = \hbox{$\frac{1}{2}$}
\epsilon_{\alpha\beta\gamma\delta} M^{\gamma\delta}$, which transform as the
second--rank complex selfdual representation of $SU(4)$,
\begin{align}
\label{6.1}
&\psi_\mu^\alpha = \begin{pmatrix} \psi_\mu^a \\
- i \epsilon_{\mu\nu} \bar{\psi}^{\nu a} \end{pmatrix},
\qquad
\bar{\eta}_\alpha = \begin{pmatrix} \bar{\eta}_a \\
i \zeta_a \end{pmatrix},
\qquad
\bar{\zeta}_\alpha = \begin{pmatrix} \bar{\zeta}_a \\
- i \eta_a \end{pmatrix},
\nonumber
\\
&M^{\alpha\beta} = \begin{pmatrix}
\epsilon^{ab} M & i M^{ba} \\
- i M^{ab} & \epsilon^{ab} \bar{M} \end{pmatrix},
\qquad
M_{\alpha\beta} = \begin{pmatrix}
\epsilon_{ab} \bar{M} & - i M_{ba} \\
i M_{ab} & \epsilon_{ab} M \end{pmatrix},
\end{align}
where from now on $\alpha$ denotes the internal group index of $SU(4)$ taking
4 values. By virtue of (\ref{4.2}) and (\ref{5.1}), it is easily seen that
the matrices $M_{\alpha\beta}$ and $M^{\alpha\beta}$ can be written in the
form
\begin{equation*}
M_{\alpha\beta} = (\gamma_m)_{\alpha\beta} M^m,
\qquad
M^{\alpha\beta} = (\gamma_m)_{\alpha\beta}^* M^m = \hbox{$\frac{1}{2}$}
\epsilon^{\alpha\beta\gamma\delta} M_{\gamma\delta},
\qquad
m = 1, \ldots, 6,
\end{equation*}
with $M^m = \{ A_{3,4,5}, V_{3,4,5} \}$, where $(\gamma_m)_{\alpha\beta}$ are
the generators of the group $SO(6)$,
\begin{equation*}
(\gamma_m)_{\alpha\beta} = ( \gamma_1 C_5, \gamma_2 C_5, \gamma_3 C_5,
\gamma_4 C_5, \gamma_5 C_5, C_5 ),
\qquad
\gamma_4 = - i \gamma_0,
\qquad
C_5 = \gamma_5 C_4;
\end{equation*}
the $SO(1,3)$ Dirac matrices $\gamma_\mu$ ($\mu = 0,1,2,3$) and
$\gamma_5$ are defined as in Section 3. Furthermore, analogous to
$M_{\alpha\beta}$, we combine the 6 auxiliary vector fields
($E_\mu$, $\bar{E}_\mu$, $E_\mu^{ab}$) to a $SU(4)$--sextet
$E_\mu^{\alpha\beta}$.

In terms of these new fields from (\ref{5.2}) one gets a $N_T = 8$
Hodge--type cohomological gauge theory with the larger global symmetry group
$SU(4)$,
\begin{align}
\label{6.2}
S^{(N_T = 8)} = \int_E d^2x\, {\rm tr} \Bigr\{&
\hbox{$\frac{1}{4}$} i \epsilon^{\mu\nu}
B F_{\mu\nu}(A + i V) - \hbox{$\frac{1}{4}$} i \epsilon^{\mu\nu}
\bar{B} F_{\mu\nu}(A - i V) - \hbox{$\frac{1}{2}$} \bar{B} B
\nonumber
\\
& - \epsilon^{\mu\nu} \bar{\zeta}_\alpha D_\mu(A + i V) \psi_\nu^\alpha -
\bar{\eta}_\alpha D^\mu(A - i V) \psi_\mu^\alpha -
\hbox{$\frac{1}{4}$} E^\mu_{\alpha\beta} E_\mu^{\alpha\beta}
\nonumber
\\
& + \hbox{$\frac{1}{2}$} i \epsilon^{\mu\nu}
M_{\alpha\beta} \{ \psi_\mu^\alpha, \psi_\nu^\beta \} +
i M^{\alpha\beta} \{ \bar{\eta}_\alpha, \bar{\zeta}_\beta \} -
Y D^\mu(A) V_\mu - \hbox{$\frac{1}{2}$} Y^2
\phantom{\frac{1}{2}}
\nonumber
\\
& + \hbox{$\frac{1}{8}$}
D^\mu(A + i V) M_{\alpha\beta} D_\mu(A - i V) M^{\alpha\beta} +
\hbox{$\frac{1}{64}$} [ M_{\alpha\beta}, M_{\gamma\delta} ]
[ M^{\alpha\beta}, M^{\gamma\delta} ] \Bigr\}.
\end{align}
In this $SU(4)$ symmetric form the Hodge--type $\star$ symmetry is now
defined by the replacements
\begin{equation}
\label{6.3}
\varphi \equiv \begin{bmatrix}
\partial_\mu & A_\mu & V_\mu &
\\
\psi_\mu^\alpha & \bar{\eta}_\alpha & \bar{\zeta}_\alpha & M^{\alpha\beta}
\\
B & \bar{B} & Y & E_\mu^{\alpha\beta}
\end{bmatrix}
\quad \Rightarrow \quad
\star \varphi = \begin{bmatrix}
\epsilon_{\mu\nu} \partial^\nu & \epsilon_{\mu\nu} A^\nu &
- \epsilon_{\mu\nu} V^\nu &
\\
- i \psi_\mu^\alpha & - i \bar{\zeta}_\alpha & i \bar{\eta}_\alpha &
- M^{\alpha\beta}
\\
- \bar{B} & - B & - Y & \epsilon_{\mu\nu} E^{\nu \alpha\beta}
\end{bmatrix}.
\end{equation}

From (\ref{5.4}) one can immediately read off that both the
$SU(2)$-- and $\overline{SU(2)}$--doublets ($Q^a$, $\bar{Q}^a$)
and ($^\star Q^a$, $^\star \bar{Q}^a$) should  fit into
$SU(4)$--quartets as follows,
\begin{equation*}
Q^\alpha = \begin{pmatrix} Q^a \\ i \,^\star \bar{Q}^a \end{pmatrix},
\qquad
^\star Q^\alpha = - P \star Q^\alpha \star =
\begin{pmatrix} \,^\star Q^a \\ i \bar{Q}^a \end{pmatrix},
\end{equation*}
both charges being interrelated by the $\star$ operation, such that the
Hodge--type cohomological superalgebra (\ref{5.4}) simplifies into
\begin{equation*}
\{ Q^\alpha, Q^\beta \} = 0,
\qquad
\{ Q^\alpha, \,^\star Q^\beta \} = - 2 \delta_G(M^{\alpha\beta}),
\qquad
\{ \,^\star Q^\alpha, \,^\star Q^\beta \} = 0.
\end{equation*}
Below, in Section 7, it will be shown that the action (\ref{6.2}) can be also
obtained from the twisted $N = 16$, $D = 2$ SYM with R--symmetry group
$SU(4) \otimes U(1)$. Since the relationship between the twisted and
untwisted supercharges is rather involved we shall now describe the complete
set of transformations which leave (\ref{6.2}) invariant. The transformation
rules for the topological shift symmetries $Q^\alpha$ are
\begin{align}
\label{6.4}
&Q^\alpha A_\mu = \psi_\mu^\alpha,
\nonumber
\\
&Q^\alpha V_\mu = - i \psi_\mu^\alpha,
\nonumber
\\
&Q^\alpha M_{\beta\gamma} = 2 i (
\delta^\alpha_{~\beta} \bar{\zeta}_\gamma -
\delta^\alpha_{~\gamma} \bar{\zeta}_\beta ),
\nonumber
\\
&Q^\alpha \psi_\mu^\beta = E_\mu^{\alpha\beta} -
i \epsilon_{\mu\nu} D^\nu(A - i V) M^{\alpha\beta},
\nonumber
\\
&Q^\alpha \bar{\zeta}_\beta = i \delta^\alpha_{~\beta} B,
\nonumber
\\
&Q^\alpha B = 0,
\nonumber
\\
&Q^\alpha \bar{\eta}_\beta = i \delta^\alpha_{~\beta} Y +
\hbox{$\frac{1}{2}$} [ M^{\alpha\gamma}, M_{\gamma\beta} ],
\nonumber
\\
&Q^\alpha Y = [ M^{\alpha\beta}, \bar{\zeta}_\beta ],
\nonumber
\\
&Q^\alpha \bar{B} = - 2 [ M^{\alpha\beta}, \bar{\eta}_\beta ],
\nonumber
\\
&Q^\alpha E^\mu_{\beta\gamma} = \delta^\alpha_{~[\beta} \bigr(
\epsilon^{\mu\nu} D_\nu(A + i V) \bar{\zeta}_{\gamma]} -
D^\mu(A - i V) \bar{\eta}_{\gamma]} -
i \epsilon^{\mu\nu} [ M_{\gamma]\delta}, \psi_\nu^\delta ] \bigr).
\end{align}
From combining $Q^\alpha$ with the above displayed Hodge--type $\star$
symmetry one gets the corresponding transformation rules for the topological
co--shift symmetries $^\star Q^\alpha$.

Furthermore, by an explicit calculation one can verify that (\ref{6.2}) is
also invariant under the following {\it on--shell} vector supersymmetries,
\begin{align}
\label{6.5}
&\bar{Q}_{\mu \alpha} A_\nu = \delta_{\mu\nu} \bar{\eta}_\alpha -
\epsilon_{\mu\nu} \bar{\zeta}_\alpha,
\nonumber
\\
&\bar{Q}_{\mu \alpha} V_\nu = - i \delta_{\mu\nu} \bar{\eta}_\alpha -
i \epsilon_{\mu\nu} \bar{\zeta}_\alpha,
\nonumber
\\
&\bar{Q}_{\mu \alpha} M^{\beta\gamma} = 2 i \epsilon_{\mu\nu} (
\delta_\alpha^{~\beta} \psi^{\nu \gamma} -
\delta_\alpha^{~\gamma} \psi^{\nu \beta} ),
\nonumber
\\
&\bar{Q}_{\mu \alpha} \bar{\zeta}_\beta = \epsilon_{\mu\nu}
E^\nu_{\alpha\beta} + i D_\mu(A - i V) M_{\alpha\beta},
\nonumber
\\
&\bar{Q}_{\mu \alpha} \bar{\eta}_\beta = E_{\mu \alpha\beta} +
i \epsilon_{\mu\nu} D^\nu(A + i V) M_{\alpha\beta},
\nonumber
\\
&\bar{Q}_{\mu \alpha} \psi_\nu^\beta = - 2 \delta_\alpha^{~\beta} F_{\mu\nu}(A) -
2 i \delta_\alpha^{~\beta} D_\mu(A) V_\nu -
i \delta_\alpha^{~\beta} \delta_{\mu\nu} Y -
i \delta_\alpha^{~\beta} \epsilon_{\mu\nu} \bar{B} +
\hbox{$\frac{1}{2}$} \delta_{\mu\nu} [ M_{\alpha\gamma}, M^{\gamma\beta} ],
\nonumber
\\
&\bar{Q}_{\mu \alpha} \bar{B} = 2 i \epsilon_{\mu\nu}
D^\nu(A + i V) \bar{\eta}_\alpha,
\nonumber
\\
&\bar{Q}_{\mu \alpha} Y = 2 i D_\mu(A - i V) \bar{\eta}_\alpha -
\epsilon_{\mu\nu} [ M_{\alpha\beta}, \psi^{\nu \beta} ],
\nonumber
\\
&\bar{Q}_{\mu \alpha} B = 2 i \epsilon_{\mu\nu} D^\nu(A - i V) \bar{\eta}_\alpha +
4 i D_\mu(A) \bar{\zeta}_\alpha + 2 [ M_{\alpha\beta}, \psi_\mu^\beta ],
\nonumber
\\
&\bar{Q}_{\mu \alpha} E_\nu^{\beta\gamma} = - \delta_\alpha^{~[\beta}
D_{[\mu}(A + i V) \psi_{\nu]}^{~\gamma]} -
\delta_\alpha^{~[\beta} \delta_{\mu\nu} D^\rho(A - i V) \psi_\rho^{\gamma]} +
i \delta_\alpha^{~[\beta} [ M^{\gamma] \delta},
\epsilon_{\mu\nu} \bar{\eta}_\delta -
\delta_{\mu\nu} \bar{\zeta}_\delta ].
\end{align}
The vector supercharges $\bar{Q}_{\mu \alpha}$, together with $Q^\alpha$
and $^\star Q^\alpha$, obey the anticommutation relations
\begin{equation*}
\{ Q^\alpha, \bar{Q}_{\mu \beta} \} \doteq
- 2 \delta^\alpha_{~\beta} (
\partial_\mu + \delta_G(A_\mu - i V_\mu) ),
\qquad
\{ \,^\star Q^\alpha, \bar{Q}_{\mu \beta} \} \doteq
- 2 \delta^\alpha_{~\beta} (
\partial_\mu + \delta_G(A_\mu + i V_\mu) ).
\end{equation*}
The action (\ref{6.2}) is also invariant under the co--vector supersymmetries
\begin{equation*}
^\star \bar{Q}_{\mu \alpha} = - P \star \bar{Q}_{\mu \alpha} \star \doteq
i \bar{Q}_{\mu \alpha},
\end{equation*}
which on--shell, i.e., by using {\it only} the equations of motion of the
auxiliarly fields, becomes $i$ times the vector supersymmetries.
Hence, it holds
\begin{equation*}
( Q^\alpha, \,^\star Q^\alpha, \bar{Q}_{\mu \alpha} ) S^{(N_T = 8)} = 0,
\end{equation*}
and the total number of (real) supercharges is actually $N = 16$. Let us
remark that we were not able to find an appropiate set of auxiliary fields
in order to complete the superalgebra of the full set of both scalar {\it and}
vector supercharges off--shell.

Finally, let us mention that both the components of the
complexified gauge field, $A_\alpha \pm i V_\alpha$, do not
possess a harmonic part, although $A_\alpha - i V_\alpha$ and
$A_\alpha + i V_\alpha$ are, respectively, $^\alpha Q$-- and
$^\star Q^\alpha$--invariant. This is owing to the fact that,
instead of the gauge field $A_\alpha$ and the co--vector field
$V_\alpha$, in the Euclidean space one can view $A_\alpha - i
V_\alpha$ and $A_\alpha + i V_\alpha$ as two independent fields
belonging to the gauge multiplet of the theory.

\bigskip
\begin{flushleft}
{\large{\bf 7. The $N_T = 8$ topological twist of $N = 16$, $D = 2$ super
Yang--Mills theory}}
\end{flushleft}
\medskip

Now, as anticipated in the previous section, we show that the $N_T = 8$
Hodge--type cohomological theory with global symmetry group $SU(4)$
arises from a topological twist of $N = 16$, $D = 2$ SYM.
A group theoretical description of this twist has been given
in \cite{3}: First, one dimensionally reduces $N = 1$, $D = 10$ SYM
to $D = 2$ by breaking down the Lorentz group according to $SO(1,9)
\supset SO(1,1) \otimes Spin(8)$, where $Spin(8)$ is the covering of the
R--symmetry group $SO(8)$ of the dimensionally reduced Minkowskian
$N = 16$, $D = 2$ SYM.
Then, one considers the branching $SO(8) \supset SU(4) \otimes U(1)$ and
performs a Wick rotation into the Euclidean space. Afterwards one twists the
Euclidean rotation group $SO_E(2) \sim U_E(1)$ in $D = 2$ by the $U(1)$ of
the R--symmetry group (by simply adding up the both $U(1)$ charges),
thereby leaving the group $SU(4)$ intact.

Here, we shall proceed in another way. Starting from Euclidean
$N = 4$, $D = 4$ SYM, which already is manifestly $SU(4)$--invariant, and
performing a dimensional reduction to $D = 2$ we get the Euclidean
$N = 16$, $D = 2$ SYM. Then, in order to reveal how this theory should be
twisted, we carry out the topological B--twist \cite{11} of the Euclidean
$N = 4$, $D = 4$ SYM as well as a dimensional reduction to $D = 2$ and compare
this twisted theory with the untwisted one and with the topological theory
(\ref{6.2}). From this comparison one immediately reads off how the Euclidean
$N = 16$, $D = 2$ SYM has to be twisted in order to get the $N_T = 8$
Hodge--type cohomological theory (\ref{6.2}).
Since the relationship between the twisted and the untwisted fields is rather
complex we describe this twisting procedure in some detail.

The field content of $N = 4$, $D = 4$ SYM consists of an anti--hermitean
gauge field $A_\mu$, a Majorana spinor $\lambda_{A \alpha}$ and its conjugate
one $\bar{\lambda}_{\dot{A}}^{\!~~\alpha}$, which
transform as the fundamental and its complex conjugate representation of
$SU(4)$, respectively, and a set of complex scalar fields $G_{\alpha\beta} =
\hbox{$\frac{1}{2}$} \epsilon_{\alpha\beta\gamma\delta} G^{\gamma\delta}$,
which transform as the second--rank complex selfdual representation
of $SU(4)$. All the fields take their values in the Lie algebra $Lie(G)$
of the gauge group $G$.

In  Euclidean space this theory has the invariant action \cite{18}
\begin{align}
\label{7.1}
S^{(N = 4)} = \int_E d^4x\, {\rm tr} \Bigr\{&
\hbox{$\frac{1}{4}$} F_{\mu\nu} F^{\mu\nu} -
i \bar{\lambda}_{\dot{A}}^{\!~~\alpha} (\sigma_\mu)^{\dot{A} B}
D^\mu \lambda_{B \alpha} +
\hbox{$\frac{1}{64}$} [ G_{\alpha\beta}, G_{\gamma\delta} ]
[ G^{\alpha\beta}, G^{\gamma\delta} ]
\nonumber
\\
& - \hbox{$\frac{1}{2}$} i \lambda_{A \alpha}
[ G^{\alpha\beta}, \lambda^A_{\!~~\beta} ] -
\hbox{$\frac{1}{2}$} i \bar{\lambda}^{\dot{A} \alpha}
[ G_{\alpha\beta}, \bar{\lambda}_{\dot{A}}^{\!~~\beta} ] +
\hbox{$\frac{1}{8}$} D_\mu G_{\alpha\beta} D^\mu G^{\alpha\beta} \Bigr\},
\end{align}
where the numerically invariant tensors $(\sigma_\mu)^{A \dot{B}}$ and
$(\sigma_\mu)_{\dot{A} B}$ are the Clebsch--Cordon coefficients relating the
$(1/2,1/2)$ representation of $SL(2,C)$ to the vector representation of
$SO(4)$,
\begin{alignat}{2}
\label{7.2}
&(\sigma_\mu)^{\dot{A} B} = ( -i \sigma_1, -i \sigma_2, -i \sigma_3, I_2 ),
&\qquad
&(\sigma_\mu)_{\dot{A} B} \equiv (\sigma_\mu)^{\dot{C} D}
\epsilon_{\dot{C}\dot{A}} \epsilon_{DB} = (\sigma_\mu^*)^{\dot{A} B},
\nonumber
\\
&(\sigma_\mu)_{A \dot{B}} = ( i \sigma_1, i \sigma_2, i \sigma_3, I_2 ),
&\qquad
&(\sigma_\mu)^{A \dot{B}} \equiv \epsilon^{AC} \epsilon^{\dot{B}\dot{D}}
(\sigma_\mu)_{C \dot{D}} = (\sigma_\mu^*)_{A \dot{B}},
\end{alignat}
$(\sigma_\mu)_{\dot{A} B}$ and $(\sigma_\mu)^{A \dot{B}}$ being the
corresponding complex conjugate coefficients. The selfdual
and anti--selfdual generators of the $SO(4)$ rotations,
$(\sigma_{\mu\nu})_{AB}$ and $(\sigma_{\mu\nu})_{\dot{A}\dot{B}}$, obey
the relations
\begin{align}
\label{7.3}
&(\sigma_\mu)^{A \dot{C}} (\sigma_\nu)_{\dot{C}}^{\!~~B} =
(\sigma_{\mu\nu})^{AB} -
\delta_{\mu\nu} \epsilon^{AB},
\nonumber
\\
&(\sigma_\rho)^{A \dot{C}} (\sigma_{\mu\nu})_{\dot{C}}^{\!~~\dot{B}} =
\delta_{\rho\mu} (\sigma_\nu)^{A \dot{B}} -
\delta_{\rho\nu} (\sigma_\mu)^{A \dot{B}} -
\epsilon_{\mu\nu\rho\sigma} (\sigma^\sigma)^{A \dot{B}},
\\
\label{7.4}
&(\sigma_\mu)_{\dot{A} C} (\sigma_\nu)^C_{\!~~\dot{B}} =
(\sigma_{\mu\nu})_{\dot{A}\dot{B}} +
\delta_{\mu\nu} \epsilon_{\dot{A}\dot{B}},
\nonumber
\\
&(\sigma_\rho)_{\dot{A} C} (\sigma_{\mu\nu})^C_{\!~~B} =
\delta_{\rho\mu} (\sigma_\nu)_{\dot{A} B} -
\delta_{\rho\nu} (\sigma_\mu)_{\dot{A} B} +
\epsilon_{\mu\nu\rho\sigma} (\sigma^\sigma)_{\dot{A} B}.
\end{align}
The spinor index $A$ (and analogous $\dot{A}$) is raised and lowered as
follows: $\epsilon^{AC} \varphi_C^{\!~~B} = \varphi^{AB}$ and
$\varphi_A^{\!~~C} \epsilon_{CB} = \varphi_{AB}$,
where $\epsilon_{AB}$ (and analogous $\epsilon_{\dot{A}\dot{B}}$) is
the invariant tensor of the group $SU(2)$, $\epsilon_{12} = \epsilon^{12} =
\epsilon_{\dot{1}\dot{2}} = \epsilon^{\dot{1}\dot{2}} = 1$.

The action (\ref{7.1}) is manifestly invariant under hermitean
conjugation:
\begin{equation*}
( A_\mu, \lambda_{A \alpha}, \bar{\lambda}^{\dot{A} \alpha},
G^{\alpha\beta} ) \rightarrow
( - A_\mu, \bar{\lambda}_{\dot{A}}^{\!~~\alpha}, \lambda^A_{\!~~\alpha},
G_{\alpha\beta} ).
\end{equation*}
Furthermore, making use of (\ref{7.3}) and (\ref{7.4}), by a brief
calculation one verifies that (\ref{7.1}) is invariant under the following
{\it on--shell} supersymmetry transformations,
\begin{align*}
&Q_A^{\!~~\alpha} A_\mu = - i (\sigma_\mu)_{A \dot{B}}
\bar{\lambda}^{\dot{B} \alpha},
\\
&Q_A^{\!~~\alpha} \bar{\lambda}_{\dot{B}}^{\!~~\beta} =
(\sigma^\mu)_{A \dot{B}} D_\mu G^{\alpha\beta},
\\
&Q_A^{\!~~\alpha} G_{\beta\gamma} =
2 i ( \delta^\alpha_{~\beta} \lambda_{A \gamma} -
\delta^\alpha_{~\gamma} \lambda_{A \beta} ),
\\
&Q_A^{\!~~\alpha} \lambda_{B \beta} =
- \hbox{$\frac{1}{2}$} \delta^\alpha_{~\beta}
(\sigma^{\mu\nu})_{AB} F_{\mu\nu} -
\hbox{$\frac{1}{2}$} \epsilon_{AB} [ G^{\alpha\gamma}, G_{\gamma\beta} ]
\end{align*}
and
\begin{align*}
&\bar{Q}_{\dot{A} \alpha} A_\mu = i (\sigma_\mu)_{\dot{A} B}
\lambda^B_{\!~~\alpha},
\\
&\bar{Q}_{\dot{A} \alpha} \lambda_{B \beta} =
(\sigma^\mu)_{\dot{A} B} D_\mu G_{\alpha\beta},
\\
&\bar{Q}_{\dot{A} \alpha} G^{\beta\gamma} =
2 i ( \delta_\alpha^{~\beta} \bar{\lambda}_{\dot{A}}^{\!~~\gamma} -
\delta_\alpha^{~\gamma} \bar{\lambda}_{\dot{A}}^{\!~~\beta} ),
\\
&\bar{Q}_{\dot{A} \alpha} \bar{\lambda}_{\dot{B}}^{\!~~\beta} =
- \hbox{$\frac{1}{2}$} \delta_\alpha^{~\beta}
(\sigma^{\mu\nu})_{\dot{A}\dot{B}} F_{\mu\nu} +
\hbox{$\frac{1}{2}$} \epsilon_{\dot{A}\dot{B}}
[ G_{\alpha\gamma}, G^{\gamma\beta} ].
\end{align*}
Let us recall that it is not possible to complete the superalgebra off--shell
with a finite number of auxiliary fields \cite{21}.

In order to perform in (\ref{7.1}) a dimensional reduction to $D = 2$ we
rename the third and fourth component of $A_\mu$ according to
\begin{equation}
\label{7.5}
A_3 = \hbox{$\frac{1}{2}$} ( \phi + \bar{\phi} ),
\qquad
A_4 = \hbox{$\frac{1}{2}$} i ( \phi - \bar{\phi} ),
\end{equation}
reserving the notation $A_\mu$ ($\mu = 1,2$) for the gauge field in $D = 2$.
Moreover, we decompose the components of $(\sigma_\mu)_{\dot{A}}^{\!~~B}$,
$(\sigma_{\mu\nu})_{\dot{A}}^{\!~~\dot{B}}$ and
$(\sigma_\mu)_A^{\!~~\dot{B}}$, $(\sigma_{\mu\nu})_A^{\!~~B}$ in the
following manner,
\begin{alignat}{2}
\label{7.6}
&(\sigma_\mu)_{\dot{A}}^{\!~~B}
\rightarrow
i (\sigma_\mu)_A^{\!~~B},
&\qquad
&(\sigma_\mu)_A^{\!~~\dot{B}}
\rightarrow
i (\sigma_\mu)_A^{\!~~B},
\nonumber
\\
&(\sigma_3)_{\dot{A}}^{\!~~B}
\rightarrow
- i (\sigma_3)_A^{\!~~B},
&\qquad
&(\sigma_3)_A^{\!~~\dot{B}}
\rightarrow
- i (\sigma_3)_A^{\!~~B},
\nonumber
\\
&(\sigma_4)_{\dot{A}}^{\!~~B}
\rightarrow
\delta_A^{\!~~B},
&\qquad
&(\sigma_4)_A^{\!~~\dot{B}}
\rightarrow
- \delta_A^{\!~~B},
\nonumber
\\
&(\sigma_{\mu\nu})_{\dot{A}}^{\!~~\dot{B}}
\rightarrow
- i \epsilon_{\mu\nu} (\sigma_3)_A^{\!~~B},
&\qquad
&(\sigma_{\mu\nu})_A^{\!~~B}
\rightarrow
i \epsilon_{\mu\nu} (\sigma_3)_A^{\!~~B},
\nonumber
\\
&(\sigma_{\mu 3})_{\dot{A}}^{\!~~\dot{B}}
\rightarrow
i \epsilon_{\mu\nu} (\sigma^\nu)_A^{\!~~B},
&\qquad
&(\sigma_{\mu 3})_A^{\!~~B}
\rightarrow
- i \epsilon_{\mu\nu} (\sigma^\nu)_A^{\!~~B},
\nonumber
\\
&(\sigma_{\mu 4})_{\dot{A}}^{\!~~\dot{B}}
\rightarrow
i (\sigma_\mu)_A^{\!~~B},
&\qquad
&(\sigma_{\mu 4})_A^{\!~~B}
\rightarrow
i (\sigma_\mu)_A^{\!~~B},
\nonumber
\\
&(\sigma_{34})_{\dot{A}}^{\!~~\dot{B}}
\rightarrow
- i (\sigma_3)_A^{\!~~B},
&\qquad
&(\sigma_{34})_A^{\!~~B}
\rightarrow
- i (\sigma_3)_A^{\!~~B},
\end{alignat}
such that  both the relations (\ref{7.3}) and (\ref{7.4}) become the
algebra of the Pauli matrices (observe that
$(\sigma_\mu, \sigma_3)_A^{\!~~B} \equiv (\sigma_1, \sigma_2, \sigma_3)$),
\begin{align*}
&(\sigma_\mu)_A^{\!~~C} (\sigma_\nu)_{CB} = \delta_{\mu\nu} \epsilon_{AB} +
i \epsilon_{\mu\nu} (\sigma_3)_{AB},
\\
&(\sigma_\mu)_A^{\!~~C} (\sigma_3)_{CB} = - i \epsilon_{\mu\nu}
(\sigma^\nu)_{AB},
\\
&(\sigma_3)_A^{\!~~C} (\sigma_3)_{CB} = \epsilon_{AB}.
\end{align*}

Then, from the action (\ref{7.1}), for the Euclidean action of the $N = 16$,
$D = 2$ SYM we obtain
\begin{align}
\label{7.7}
S^{(N = 16)} = \int_E d^2x\, {\rm tr} \Bigr\{&
\hbox{$\frac{1}{4}$} F_{\mu\nu} F^{\mu\nu} +
\hbox{$\frac{1}{2}$} D_\mu \bar{\phi} D^\mu \phi -
\hbox{$\frac{1}{8}$} [ \bar{\phi}, \phi ]^2
\nonumber
\\
& - \hbox{$\frac{1}{2}$} \bar{\lambda}_A^{\!~~\alpha} (\sigma_3)^{AB}
[ \phi + \bar{\phi}, \lambda_{B \alpha} ] +
\hbox{\large$\frac{1}{2}$} \bar{\lambda}^{A \alpha}
[ \phi - \bar{\phi}, \lambda_{A \alpha} ]
\nonumber
\\
& + \bar{\lambda}_A^{\!~~\alpha} (\sigma_\mu)^{AB}
D^\mu \lambda_{B \alpha} -
\hbox{$\frac{1}{2}$} i \lambda_{A \alpha}
[ G^{\alpha\beta}, \lambda^A_{\!~~\beta} ] -
\hbox{$\frac{1}{2}$} i \bar{\lambda}^{A \alpha}
[ G_{\alpha\beta}, \bar{\lambda}_A^{\!~~\beta} ]
\phantom{\frac{1}{2}}
\nonumber
\\
& + \hbox{$\frac{1}{8}$} D_\mu G_{\alpha\beta} D^\mu G^{\alpha\beta} +
\hbox{$\frac{1}{8}$} [ \bar{\phi}, G_{\alpha\beta} ] [ \phi, G^{\alpha\beta} ] +
\hbox{$\frac{1}{64}$} [ G_{\alpha\beta}, G_{\gamma\delta} ]
[ G^{\alpha\beta}, G^{\gamma\delta} ] \Bigr\}.
\end{align}
Since the decompositions (\ref{7.6}) include some factors of $i$,
the action (\ref{7.7}) is no longer manifestly invariant under hermitean
conjugation. Rather, it is invariant under the following $Z_2$ symmetry,
\begin{equation}
\label{7.8}
Z_2:
\qquad
( A_\mu, \phi, \bar{\phi},
\lambda_{A \alpha}, \bar{\lambda}^{A \alpha}, G^{\alpha\beta} ) \rightarrow
( A_\mu, \bar{\phi}, \phi,
- \bar{\lambda}_A^{\!~~\alpha}, - \lambda^A_{\!~~\alpha}, G_{\alpha\beta} ).
\end{equation}
If we denote the $N = 16$ spinorial supercharges in $D = 2$ with
$Q_A^{\!~~\alpha}$ and $\bar{Q}_{A \alpha}$, which are interchanged by the
$Z_2$ symmetry (\ref{7.8}), the transformation rules for $Q_A^{\!~~\alpha}$
are
\begin{align}
\label{7.9}
&Q_A^{\!~~\alpha} A_\mu = (\sigma_\mu)_{AB} \bar{\lambda}^{B \alpha},
\nonumber
\\
&Q_A^{\!~~\alpha} \phi = - (\sigma_3)_{AB} \bar{\lambda}^{B \alpha} -
\bar{\lambda}_A^{\!~~\alpha},
\nonumber
\\
&Q_A^{\!~~\alpha} \bar{\phi} = - (\sigma_3)_{AB} \bar{\lambda}^{B \alpha} +
\bar{\lambda}_A^{\!~~\alpha},
\nonumber
\\
&Q_A^{\!~~\alpha} G_{\beta\gamma} =
2 i ( \delta^\alpha_{~\beta} \lambda_{A \gamma} -
\delta^\alpha_{~\gamma} \lambda_{A \beta} ),
\nonumber
\\
&Q_A^{\!~~\alpha} \bar{\lambda}_B^{\!~~\beta} =
\hbox{$\frac{1}{2}$} i (\sigma^\mu)_{AB} D_\mu G^{\alpha\beta} -
\hbox{$\frac{1}{2}$} i (\sigma_3)_{AB} [ \phi + \bar{\phi}, G^{\alpha\beta} ] -
\hbox{$\frac{1}{2}$} i \epsilon_{AB} [ \phi - \bar{\phi}, G^{\alpha\beta} ],
\nonumber
\\
&Q_A^{\!~~\alpha} \lambda_{B \beta} =
\hbox{$\frac{1}{2}$} i \delta^\alpha_{~\beta}
\epsilon^{\mu\nu} (\sigma_\nu)_{AB} D_\mu ( \phi + \bar{\phi} ) +
\hbox{$\frac{1}{2}$} \delta^\alpha_{~\beta}
(\sigma^\mu)_{AB} D_\mu ( \phi - \bar{\phi} )
\nonumber
\\
&\phantom{Q_A^{\!~~\alpha} \lambda_{B \beta} =}
+ \hbox{\large$\frac{1}{2}$} \delta^\alpha_{~\beta}
(\sigma^3)_{AB} [ \phi, \bar{\phi} ] - \hbox{$\frac{1}{2}$} i \delta^\alpha_{~\beta}
\epsilon^{\mu\nu} (\sigma_3)_{AB} F_{\mu\nu} -
\hbox{$\frac{1}{2}$} \epsilon_{AB} [ G^{\alpha\gamma}, G_{\gamma\beta} ].
\end{align}

Let us now derive the relationship between the action (\ref{7.7}) and the
cohomological theory (\ref{6.2}). Its complexity stems from a
nontrivial mixing of the scalar and vector supercharges
$Q^\alpha$, $^\star Q^\alpha$, $\bar{Q}_{\mu \alpha}$ and the spinorial
supercharges $Q^{A \alpha}$ and $\bar{Q}_{A \alpha}$, namely, the internal
indices $\alpha$ in both theories cannot be simply identified with each
other. For that reason, as explained above, we proceed as follows: First
of all, we perform in (\ref{7.1}) a topological B--twist \cite{11}, i.e., we
break the group $SU(4)$ down to $SU(2) \otimes \overline{SU(2)} \otimes U(1)$
and identify the components $\alpha = (1,2)$ and $\alpha = (3,4)$ with the
spinor indices $B = (1,2)$ and $\dot{B} = (\dot{1},\dot{2})$ of both
subgroups of the Euclidean rotation group $SO(4) = SU(2)_L \otimes SU(2)_R$,
respectively. Then, we replace in (\ref{7.1}) the fields
$\lambda_{A \alpha}$, $\bar{\lambda}^{\dot{A} \alpha}$ and $G_{\alpha\beta}$
by the twisted fields $\eta$, $\tilde{\eta}$, $\psi_\mu$, $\tilde{\psi}_\mu$,
$\chi_{\mu\nu}$ and $G$, $\bar{G}$, $V_\mu$ of the B--model according to
\begin{align}
\label{7.10}
&\lambda_{A \alpha} = \hbox{$\frac{1}{2}$} \begin{pmatrix}
\epsilon_{AB} ( \eta + \tilde{\eta} ) -
\frac{1}{4} (\sigma^{\mu\nu})_{AB} (
\chi_{\mu\nu} + \tilde{\chi}_{\mu\nu} ) \\
\phantom{-}(\sigma^\mu)_{A \dot{B}} ( \psi_\mu - \tilde{\psi}_\mu )
\end{pmatrix},
\nonumber
\\
&\bar{\lambda}^{\dot{A} \alpha} = \hbox{$\frac{1}{2}$} \begin{pmatrix}
- (\sigma^\mu)^{\dot{A} B} ( \tilde{\psi}_\mu + \psi_\mu) \\
\epsilon^{\dot{A} \dot{B}} ( \tilde{\eta} - \eta ) -
\frac{1}{4} (\sigma^{\mu\nu})^{\dot{A} \dot{B}} (
\tilde{\chi}_{\mu\nu} - \chi_{\mu\nu} )
\end{pmatrix},
\\
\intertext{and}
\label{7.11}
&G_{\alpha\beta} = \begin{pmatrix}
\epsilon_{AB} \bar{G} & i (\sigma^\mu)_{A \dot{B}} V_\mu \\
- i (\sigma^\mu)_{\dot{A} B} V_\mu & \epsilon_{\dot{A}\dot{B}} G
\end{pmatrix},
\nonumber
\\
&G^{\alpha\beta} = \begin{pmatrix}
\epsilon^{AB} G & - i (\sigma^\mu)^{A \dot{B}} V_\mu \\
i (\sigma^\mu)^{\dot{A} B} V_\mu & \epsilon^{\dot{A}\dot{B}} \bar{G}
\end{pmatrix}.
\end{align}
By using the explicit form (\ref{7.2}) of the Clebsch--Gordon coefficients
one establishes that $G_{\alpha\beta}$ and $G^{\alpha\beta}$ in (\ref{7.11})
are actually dual to each other, $G_{\alpha\beta} = \hbox{$\frac{1}{2}$}
\epsilon_{\alpha\beta\gamma\delta} G^{\gamma\delta}$. In that way,
by making use of the relations (\ref{7.3}) and (\ref{7.4}), one gets
precisely the action (\ref{4.1}) of the B--model.

As a next step, we perform in (\ref{7.10}) and (\ref{7.11}) the
decompositions (\ref{7.6}) and compare directly the resulting twisted action
with (\ref{6.2}) --- after the elimination of the auxiliary fields
$B$, $\bar{B}$, $E_\mu^{\alpha\beta}$ and $Y$ through their equations of
motion. From this comparison one can deduce that, indeed, there is a unique
relationship between each component of the twisted fields which enter into
(\ref{7.10}) and (\ref{7.11}), and the complex scalar fields $\phi$ and
$\bar{\phi}$ introduced in (\ref{7.5}), as well as the whole set of fields
$\psi_\mu^\alpha$, $\bar{\eta}_\alpha$, $\bar{\zeta}_\alpha$ and
$M_{\alpha\beta}$ which enter into (\ref{6.2}).

Collecting these results, after a lengthly calculation one obtains the
following relationships:
\begin{align}
\label{7.12}
&\lambda_{A \alpha} = \hbox{$\frac{1}{2}$} \begin{pmatrix}
i (\sigma^\mu)_{AB} ( \psi_\mu^1 - \epsilon_{\mu\nu} \psi^{\nu 3} ) +
(\sigma_3)_{AB} ( \bar{\eta}_4 + \bar{\zeta}_2 ) +
i \epsilon_{AB} ( \bar{\zeta}_4 - \bar{\eta}_2 ) \\
i (\sigma^\mu)_{AB} ( \psi_\mu^2 + \epsilon_{\mu\nu} \psi^{\nu 4} ) +
(\sigma_3)_{AB} ( \bar{\eta}_3 - \bar{\zeta}_1 ) +
i \epsilon_{AB} ( \bar{\zeta}_3 + \bar{\eta}_1 ) \end{pmatrix},
\nonumber
\\
&\bar{\lambda}^{A \alpha} = \hbox{$\frac{1}{2}$} \begin{pmatrix}
i (\sigma^\mu)^{AB} ( \epsilon_{\mu\nu} \psi^{\nu 4} - \psi_\mu^2 ) +
(\sigma_3)^{AB} ( \bar{\zeta}_1 + \bar{\eta}_3 ) -
i \epsilon^{AB} ( \bar{\eta}_1 - \bar{\zeta}_3 ) \\
i (\sigma^\mu)^{AB} ( \epsilon_{\mu\nu} \psi^{\nu 3} + \psi_\mu^1 ) +
(\sigma_3)^{AB} ( \bar{\zeta}_2 - \bar{\eta}_4 ) -
i \epsilon^{AB} ( \bar{\eta}_2 + \bar{\zeta}_4 ) \end{pmatrix},
\\
\intertext{between $\lambda_{A \alpha}$, $\bar{\lambda}^{A \alpha}$ and the
twisted spinor fields $\psi_\mu^\alpha$, $\bar{\eta}_\alpha$,
$\bar{\zeta}_\alpha$, and}
\label{7.13}
&\phi = A_3 - i A_4,
\qquad
\bar{\phi} = A_3 + i A_4,
\nonumber
\\
&G_{\alpha\beta} = \begin{pmatrix}
\epsilon_{AB} \bar{G} & - (\sigma^\mu)_{AB} V_\mu +
(\sigma_3)_{AB} V_3 - i \epsilon_{AB} V_4 \\
(\sigma^\mu)_{AB} V_\mu - (\sigma_3)_{AB} V_3 -
i \epsilon_{AB} V_4 & \epsilon_{AB} G
\end{pmatrix},
\nonumber
\\
&G^{\alpha\beta} = \begin{pmatrix}
\epsilon^{AB} G & (\sigma^\mu)^{AB} V_\mu -
(\sigma_3)^{AB} V_3 + i \epsilon^{AB} V_4 \\
- (\sigma^\mu)^{AB} V_\mu + (\sigma_3)^{AB} V_3 +
i \epsilon^{AB} V_4 & \epsilon^{AB} \bar{G}
\end{pmatrix},
\end{align}
where
\begin{alignat}{4}
\label{7.14}
&A_3 = \hbox{$\frac{1}{2}$} ( M^{12} + M^{34} ),
&\qquad
&V_3 = \hbox{$\frac{1}{2}$} i ( M^{12} - M^{34} ),
&
\qquad
&G = M^{24},
\nonumber
\\
&A_4 = \hbox{$\frac{1}{2}$} ( M^{14} + M^{23} ),
&\qquad
&V_4 = \hbox{$\frac{1}{2}$} i ( M^{14} - M^{23} ),
&\qquad
&\bar{G} = M^{31},
\end{alignat}
between $\phi$, $\bar{\phi}$, $G_{\alpha\beta}$ and the twisted vector and
scalar fields $V_\mu$ and $M_{\alpha\beta}$, respectively.

Thereby, the assigment between the spinor indices $(A,B)$ and the group
indices $(\alpha,\beta)$ is similar as before, e.g., in (\ref{7.13}) the
spinor indices $A = 1,2$ (resp. $B = 1,2$) at the upper and lower
raw (resp. at the left and right column) of the both matrices correspond
to the values $\alpha = 1,2$ and $\alpha = 3,4$ (resp. $\beta = 1,2$ and
$\beta = 3,4$) of the scalar fields $G_{\alpha\beta}$, respectively.
Let us also notice, that the relationships (\ref{7.14}) agree precisely
with the definition (\ref{6.1}) of the matrix $M^{\alpha\beta}$
(c.f., Eqs. (\ref{4.2}) and (\ref{5.1})).

As an additional check, inserting (\ref{7.12}) -- (\ref{7.14}) into (\ref{7.7}),
after a tedious calculation one verifies that the resulting twisted
action actually agrees with the topological action (\ref{6.2}).

The relationship between the spinorial supercharges $Q^{A \alpha}$ and
$\bar{Q}_{A \alpha}$, being interrelated by the replacements (\ref{7.8}),
and the twisted scalar and vector supercharges $Q^\alpha$,
$\bar{Q}_{\mu \alpha}$ and $^\star Q^\alpha$, $^\star \bar{Q}_{\mu \alpha}$,
being interchanged by the $\star$ operation (\ref{6.3}), is quite similar to
the ones of the spinor fields, Eq. (\ref{7.12}), namely
\begin{align*}
&Q^{A \alpha} = \hbox{$\frac{1}{2}$} \begin{pmatrix}
i (\sigma^\mu)^{AB} ( \bar{Q}_{\mu 1} - \epsilon_{\mu\nu} \bar{Q}^\nu_3 ) -
(\sigma_3)^{AB} ( Q^4 - i \,^\star Q^2 ) -
\epsilon^{AB} ( \,^\star Q^4 - i Q^2 ) \\
i (\sigma^\mu)^{AB} ( \bar{Q}_{\mu 2} + \epsilon_{\mu\nu} \bar{Q}^\nu_4 ) -
(\sigma_3)^{AB} ( Q^3 + i \,^\star Q^1 ) -
\epsilon^{AB} ( \,^\star Q^3 + i Q^1) \end{pmatrix},
\\
&\bar{Q}_{A \alpha} = \hbox{$\frac{1}{2}$} \begin{pmatrix}
i (\sigma^\mu)_{AB} ( \epsilon_{\mu\nu} \bar{Q}^\nu_4 - \bar{Q}_{\mu 2} ) +
(\sigma_3)_{AB} ( i \,^\star Q^1 - Q^3 ) +
\epsilon_{AB} ( i Q^1 - \,^\star Q^3 ) \\
i (\sigma^\mu)_{AB} ( \epsilon_{\mu\nu} \bar{Q}^\nu_3 + \bar{Q}_{\mu 1} ) +
(\sigma_3)_{AB} ( i \,^\star Q^2 + Q^4 ) +
\epsilon_{AB} ( i Q^2 + \,^\star Q^4 ) \end{pmatrix},
\end{align*}
i.e., after the twisting the $Z_2$ symmetry (\ref{7.8}) changes into the
Hodge--type $\star$ symmetry (\ref{6.3}).

Finally, let us mention that there is also a $N_T = 4$ topological twist of
$N = 16$, $D = 2$ SYM with global symmetry group $SO(4) \otimes SU(2)$.
This topological theory can be regarded as the $N_T = 4$ super--BF theory
coupled to a spinorial hypermultiplet. Another way of obtaining the action
of this theory is to dimensionally reduce either the higher dimensional
analogue of the Donaldson--Witten theory in $D = 8$ \cite{8,9} to $D = 2$ or
to dimensionally reduce the $N_T = 1$ half--twisted theory \cite{17} in
$D = 4$ to $D = 2$. However, that topological twist does not lead to another
Hodge--type cohomological theory, since the underlying cohomology is
equivariantly nilpotent and not strictly nilpotent as it should be.

\bigskip
\begin{flushleft}
{\large{\bf Concluding remarks}}
\end{flushleft}
\medskip

In this paper we have given a further example of a Hodge--type cohomological
gauge theory in $D = 2$, but this time with maximal number of $N_T = 8$
topological supercharges and largest global symmetry group $SU(4)$.
This topological theory can be obtained either by a $N_T = 1$ topological
twist of the $N = 2$, $D = 5$ SYM with R--symmetry group $SO(5)$ and
performing afterwards a ordinary dimensional reduction to $D = 2$ or by a
$N_T = 8$ topological twist of the $N = 16$, $D = 2$ SYM with
R--symmetry group $SU(4) \otimes U(1)$. This example gives rise to the
conjecture that, in general, the gauge--fixing procedure based on harmonic
gauges seems to be possible only for very constrained gauge systems with
$N_T \geq 4$ scalar supercharges.

So far, within our model we have taken into account only the
equivariant part of the BRST complex, i.e., the part associated
with the shift and co--shift symmetries $Q^\alpha$ and $^\star
Q^\alpha$. But, it should be emphasized that the $Q^\alpha$-- and
$^\star Q^\alpha$--cohomology, although empty in the space of
unrestricted local functionals of the fields, becomes nonempty if
gauge invariance is imposed on these functionals \cite{22}.
Therefore, it is important to prove whether or not for the basic
cohomology, i.e., the BRST complex including also the ordinary
gauge symmetry $\delta_G(C)$, the underlying Hodge--structure of
the theory can be preserved as well. This, of course, is a
nontrivial taks and will be the subject of a subsequent paper.

In that paper, we have only found examples in $D = 2$, where the gauge
fields have no local dynamics and where the Hodge--type $\star$ symmetry can
be rather simply implemented into the theory. In order to get some intuition
about the general structure of Hodge--type cohomological theories it would
be surely very useful to find other examples in higher dimensions.

We have conjectured that the higher dimensional analogue of the Blau--Thompson
model is the only topological one which can be constructed in $D = 5$.
In order to prove that, it is necessary to go into a closer analysis of the
structure of all the other cohomological gauge theories in $D = 5$.


\bigskip
\noindent
{\bf Acknowledgement} One of the authors (B.G.) kindly acknowledges the
warm hospitality of the Institute of Physics, Sao Paulo University,
where this work was finished, and the
financial support of this stay by DAAD -- FAPESP scientific exchange program.
He also expresses his sincere thanks to
D. Gitman for various enlightening discussions.


\end{document}